\documentclass[twocolumn,aps,pra,longbibliography,noeprint,nourl,superscriptaddress,amsmath,amssymb]{revtex4-1}

\usepackage{graphicx}
\usepackage{color}
\usepackage{upgreek}
\usepackage{float}
\usepackage{amsmath,amssymb}

\begin{document}

\title{Imaging domain reversal in an ultrathin van der Waals ferromagnet}

\author{David A. Broadway}
\thanks{These authors contributed equally to this work.}
\affiliation{School of Physics, University of Melbourne, Parkville, VIC 3010, Australia}
\affiliation{Centre for Quantum Computation and Communication Technology, School of Physics, University of Melbourne, Parkville, VIC 3010, Australia}
\address{Current address: Department of Physics, University of Basel, Klingelbergstrasse 82, Basel CH-4056, Switzerland}

\author{Sam C. Scholten}
\thanks{These authors contributed equally to this work.}
\affiliation{School of Physics, University of Melbourne, Parkville, VIC 3010, Australia}
\affiliation{Centre for Quantum Computation and Communication Technology, School of Physics, University of Melbourne, Parkville, VIC 3010, Australia}

\author{Cheng Tan}
\thanks{These authors contributed equally to this work.}
\affiliation{School of Science, RMIT University, Melbourne, VIC 3000, Australia}

\author{Nikolai Dontschuk}
\affiliation{School of Physics, University of Melbourne, Parkville, VIC 3010, Australia}
\affiliation{Centre for Quantum Computation and Communication Technology, School of Physics, University of Melbourne, Parkville, VIC 3010, Australia}

\author{Scott E. Lillie}
\affiliation{School of Physics, University of Melbourne, Parkville, VIC 3010, Australia}
\affiliation{Centre for Quantum Computation and Communication Technology, School of Physics, University of Melbourne, Parkville, VIC 3010, Australia}

\author{Brett C. Johnson}
\affiliation{School of Physics, University of Melbourne, Parkville, VIC 3010, Australia}
\affiliation{Centre for Quantum Computation and Communication Technology, School of Physics, University of Melbourne, Parkville, VIC 3010, Australia}

\author{Guolin Zheng}
\affiliation{School of Science, RMIT University, Melbourne, VIC 3000, Australia}


\author{Zhenhai Wang}
\affiliation{Skolkovo Institute of Science and Technology, Skolkovo Innovation Center, 3 Nobel Street, Moscow 143026, Russia}
\affiliation{School of Telecommunication and Information Engineering, Nanjing University of Posts and Telecommunications, Nanjing, Jiangsu 210003, China}

\author{Artem R. Oganov}
\affiliation{Skolkovo Institute of Science and Technology, Skolkovo Innovation Center, 3 Nobel Street, Moscow 143026, Russia}
\affiliation{Moscow Institute of Physics and Technology, 9 Institutsky Lane, Dolgoprudny, Moscow Region 141700, Russia}
\affiliation{International Center for Materials Discovery, Northwestern Polytechnical University, Xi'an 710072, China}

\author{Shangjie Tian}
\affiliation{Department of Physics and Beijing Key Laboratory of Optoelectronic Functional Materials \& Micro-Nano Devices, Renmin University of China, 100872 Beijing, China}

\author{Chenghe Li}
\affiliation{Department of Physics and Beijing Key Laboratory of Optoelectronic Functional Materials \& Micro-Nano Devices, Renmin University of China, 100872 Beijing, China}

\author{Hechang Lei}
\email{hlei@ruc.edu.cn}
\affiliation{Department of Physics and Beijing Key Laboratory of Optoelectronic Functional Materials \& Micro-Nano Devices, Renmin University of China, 100872 Beijing, China}

\author{Lan Wang}
\email{lan.wang@rmit.edu.au}
\affiliation{School of Science, RMIT University, Melbourne, VIC 3000, Australia}

\author{Lloyd C. L. Hollenberg}
\email{lloydch@unimelb.edu.au}
\affiliation{School of Physics, University of Melbourne, Parkville, VIC 3010, Australia}
\affiliation{Centre for Quantum Computation and Communication Technology, School of Physics, University of Melbourne, Parkville, VIC 3010, Australia}	

\author{Jean-Philippe Tetienne}
\email{jtetienne@unimelb.edu.au}
\affiliation{School of Physics, University of Melbourne, Parkville, VIC 3010, Australia}
\affiliation{Centre for Quantum Computation and Communication Technology, School of Physics, University of Melbourne, Parkville, VIC 3010, Australia}

\begin{abstract}

The recent isolation of two-dimensional van der Waals magnetic materials has uncovered rich physics that often differs from the magnetic behaviour of their bulk counterparts. However, the microscopic details of fundamental processes such as the initial magnetization or domain reversal, which govern the magnetic hysteresis, remain largely unknown in the ultrathin limit. Here we employ a widefield nitrogen-vacancy (NV) microscope to directly image these processes in few-layer flakes of magnetic semiconductor vanadium triiodide (VI$_3$). We observe complete and abrupt switching of most flakes at fields $H_c\approx0.5-1$~T (at 5 K) independent of thickness down to two atomic layers, with no intermediate partially-reversed state. The coercive field decreases as the temperature approaches the Curie temperature ($T_c\approx50$~K), however, the switching remains abrupt. We then image the initial magnetization process, which reveals thickness-dependent domain wall depinning fields well below $H_c$. These results point to ultrathin VI$_3$ being a nucleation-type hard ferromagnet, where the coercive field is set by the anisotropy-limited domain wall nucleation field. This work illustrates the power of widefield NV microscopy to investigate magnetization processes in van der Waals ferromagnets, which could be used to elucidate the origin of the hard ferromagnetic properties of other materials and explore field- and current-driven domain wall dynamics.

\end{abstract} 

\maketitle 

Two-dimensional (2D) van der Waals materials exhibiting intrinsic magnetic order have attracted enormous interest in the last few years~\cite{Huang2017,Gong2017,Burch2018,Gong2019,Gibertini2019}. However, despite much progress in the control of their magnetic properties, for example through electrostatic gating or control of the stacking order~\cite{Jiang2018,Huang2018,Li2019,Chen2019}, little is known about the mechanisms governing fundamental magnetic processes in the ultrathin limit. For instance, the extensively studied materials CrI$_3$ (a semiconductor) and Fe$_2$GeTe$_3$ (a metal) are soft ferromagnets in the bulk crystal form with a remanent magnetization far below the saturation magnetization (a few percent)~\cite{McGuire2015,Chen2013}, but surprisingly they become hard ferromagnets when exfoliated to a few atomic layers, with a near square-shaped hysteresis and a large coercive field of $H_c\sim0.1-1$~T~\cite{Huang2017,Tan2018,Fei2018,Deng2018}. Since hard ferromagnetic properties are crucial to applications, especially as a building block for van der Waals magnetic heterostructures, it is of paramount importance to understand the mechanisms that govern magnetization reversal in these systems. Unfortunately, the heterogeneous nature of exfoliated van der Waals samples precludes performing the macroscopic magnetization measurements that are normally employed to analyse bulk magnets, calling for the development of innovative approaches.     

\begin{figure}[bt!]
	\includegraphics[width=0.5\textwidth]{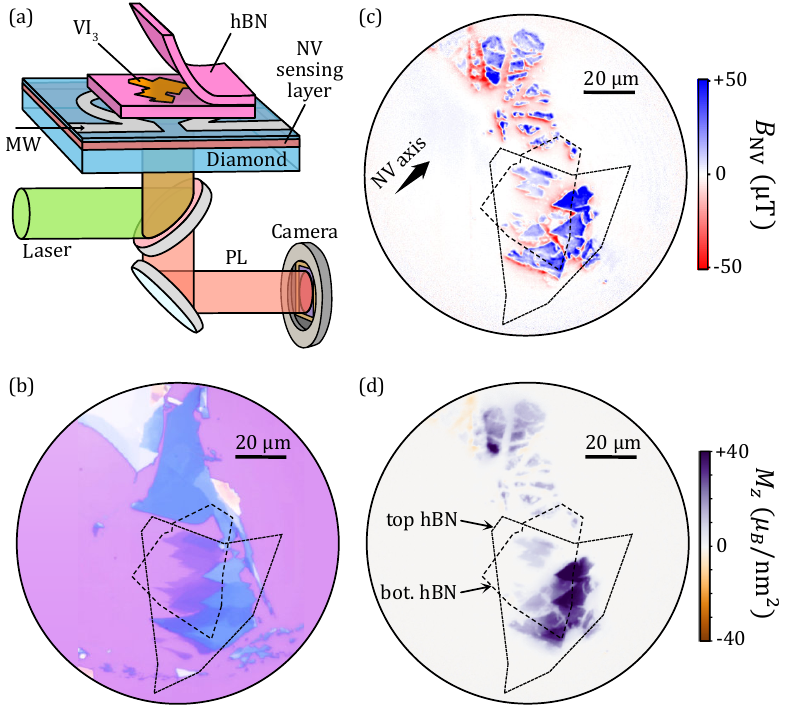}
	\caption{\textbf{Widefield NV imaging of ultrathin van der Waals magnets.} (a) Schematic of the widefield NV microscope, comprising an NV-diamond sensing chip on which the hBN/VI$_3$ heterostructures are prepared. PL: photoluminescence; MW: microwaves. (b) Optical micrograph of exfoliated VI$_3$ flakes (sample \#1) on a Si substrate prior to encapsulation and transfer to the diamond. (c) NV magnetic field map ($B_{\rm NV}$) of the flakes seen in (b) after transfer to the diamond, at 5 K. The NV projection axis is indicated by the thick arrow and points partly out of the plane. A bias field of $B_{\rm NV}^{\rm bias}=5$~mT was applied along the NV axis during the measurement, which was subtracted in the plotted map. (d) Map of the out-of-plane magnetization ($M_z$) deduced from (c). In (b-d), the dashed lines indicate the location of the hBN top and bottom layers.  
	}  
	\label{Fig1}
\end{figure}

We address this problem by directly imaging the evolution of the magnetization of ultrathin flakes using a widefield nitrogen-vacancy (NV) microscope. This recently developed magnetic imaging tool~\cite{LeSage2013,Tetienne2017,Casola2018,Lillie2020} is particularly well suited to the rapid analysis of multiple micrometre-sized samples such as exfoliated van der Waals materials, and allows us to track the domain structure of individual flakes with sub-micron spatial resolution. Our widefield NV microscope employs a diamond substrate incorporating a near-surface layer of magnetically sensitive NV centres, on which the samples are prepared (Fig.~\ref{Fig1}a). The NV layer is excited by a laser and its photoluminescence is imaged on a camera. Magnetic imaging is then realised by sweeping the frequency of an applied microwave field to obtain an optically-detected magnetic resonance spectrum. The diamond-sample assembly is placed in a cryostat allowing measurements from 4-300 K~\cite{Lillie2020}.

We studied ultrathin samples of magnetic semiconductor vanadium triiodide (VI$_3$). The magnetic properties of this van der Waals material were recently analysed in the bulk form~\cite{Tian2019,Son2019,Kong2019}, reporting an out-of-plane anisotropy and a high coercive field at low temperatures, $H_c\approx1$~T. Flakes mechanically exfoliated from a bulk VI$_3$ crystal were encapsulated with hexagonal boron nitride (hBN) to prevent degradation, and transferred to the diamond substrate. Figure~\ref{Fig1}b shows the optical image of a VI$_3$ sample on a Si substrate (prior to transfer), comprising flakes of various thicknesses from tens of nanometers down to three atomic layers in this case (based on the optical contrast, see below). The NV magnetic field image ($B_{\rm NV}$) of the same sample after transfer and cooling to a temperature of 5 K is shown in Fig.~\ref{Fig1}c, revealing magnetic signals of tens of microteslas. Knowing the projection axis of the measurement and the direction of the magnetization in the sample (out-of-plane, $z$ axis), we can reconstruct the magnetization map ($M_z$)~\cite{Thiel2019}, shown in Fig.~\ref{Fig1}d. The magnetization is given per unit surface area and reaches $50~\mu_B/{\rm nm}^2$ for the thickest flakes in this sample ($\approx20$~nm). Interestingly, the widefield image in Fig.~\ref{Fig1}d allows us to directly compare samples fully encapsulated with hBN to samples with hBN on one side only or with no hBN at all. We find that the VI$_3$ flakes that are not covered by hBN (top of the images) are still magnetic but appear fragmented with a reduced magnetization as a result of a short exposure to air during loading ($\approx5$~minutes). However, there is no visible effect of the hBN underlayer and no discontinuity in $M_z$ for flakes overlapping an hBN edge. This indicates that the magnetic properties of the flakes imaged are not measurably affected by interactions with the substrate.          

\begin{figure*}[tb!]
	\includegraphics[width=1\textwidth]{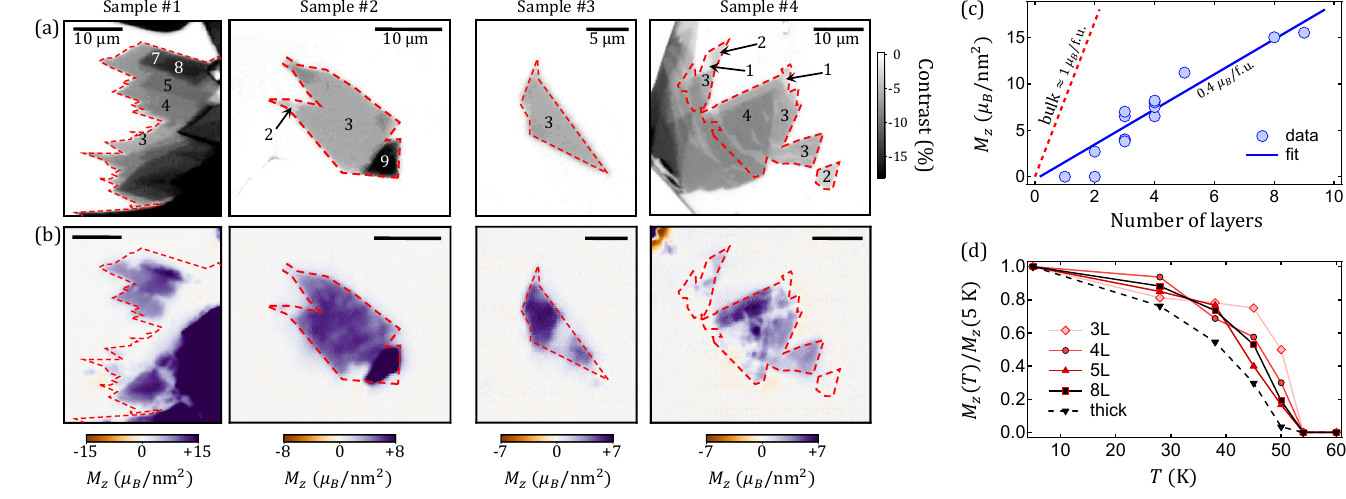}
	\caption{\textbf{Magnetization maps of few-layer VI$_3$ flakes.} (a) Optical contrast maps of four different samples prior to transfer to the diamond. The contrast is defined from the red channel intensity, relative to the background. The numbers indicate the number of atomic layers inferred from the optical contrast, corroborated by atomic force microscopy measurements (see SI, section III). The red dashed boxes indicate the flakes that were effectively transferred to the diamond (see SI, section II). (b) Corresponding magnetization maps at 5 K. (c) Maximum $M_z$ value extracted from (b) for a selection of flakes, as a function of thickness. The blue line is a fit to the data points, excluding the flakes showing zero magnetization. The red dashed line corresponds to the bulk spontaneous magnetization of about one Bohr magneton per formula unit ($1~\mu_B/$f.u.) or $5~\mu_B/{\rm nm}^2$ per layer~\cite{Son2019,Kong2019}. (d) Magnetization as a function of temperature for several flakes in sample \#1, with thickness from 3 layers (`3L') up to $\approx20$~nm (`thick').   
	}  
	\label{Fig2}
\end{figure*}

To investigate the properties of ultrathin VI$_3$, we prepared several few-layer samples (fully encapsulated with hBN) and imaged their magnetization in similar conditions. Prior to imaging, a magnetic field $B_z=+1$~T was applied in the $+z$ direction to remove the domain structure. Optical contrast maps of four samples studied are shown in Fig.~\ref{Fig2}a, from which the thickness of each flake was inferred (see SI, section III), and are compared to the corresponding magnetization maps (Fig.~\ref{Fig2}b). Most flakes down to 3 layers, as well as one bilayer flake in sample \#4, show a clear magnetic signal, demonstrating that VI$_3$ remains ferromagnetic down to two atomic layers. The absence of signal in some regions (especially in samples \#1 and \#4) is attributed to imperfect encapsulation or degradation during preparation. We also did not detect any signal from monolayer flakes (see for example the monolayer in sample \#4 adjacent to the magnetic bilayer). 

The spontaneous (areal) magnetization is estimated by taking the maximum $M_z$ value observed in the images for each domain observed. This is plotted as a function of thickness up to 9 layers in Fig.~\ref{Fig2}c, revealing a roughly linear relationship with a slope of $1.9(2)~\mu_B/{\rm nm}^2$ per atomic layer, which amounts to about $0.4~\mu_B$ per formula unit ($\mu_B/$f.u). This is somewhat lower than the spontaneous magnetization of $\approx1~\mu_B/$f.u. measured for bulk VI$_3$ crystals~\cite{Son2019,Kong2019}, again possibly due to degradation of our ultrathin VI$_3$ samples. Another possible explanation is a mixture of ferromagnetic and antiferromagnetic interlayer couplings, which would reduce the net magnetization. Indeed, our ab initio calculations predict that the antiferromagnetic state is energetically favourable in bilayer and trilayer VI$_3$ (see SI, section XIII). 

By recording magnetic field images at various temperatures $T$ (see Fig. S10), it is possible to determine the $M_z-T$ relationship as a function of thickness. This is shown in Fig.~\ref{Fig2}d, which reveals a Curie temperature of $T_c\approx50$~K similar for all the flakes analysed (down to 3 layers in this case), in agreement with the $T_c$ of bulk VI$_3$~\cite{Tian2019,Son2019,Kong2019}. The magnetization of the thickest flakes (20 nm) tends to drop more rapidly than for thinner flakes, which we attribute to domain formation in this small applied magnetic field.                    

\begin{figure*}[t!]
	\includegraphics[width=1\textwidth]{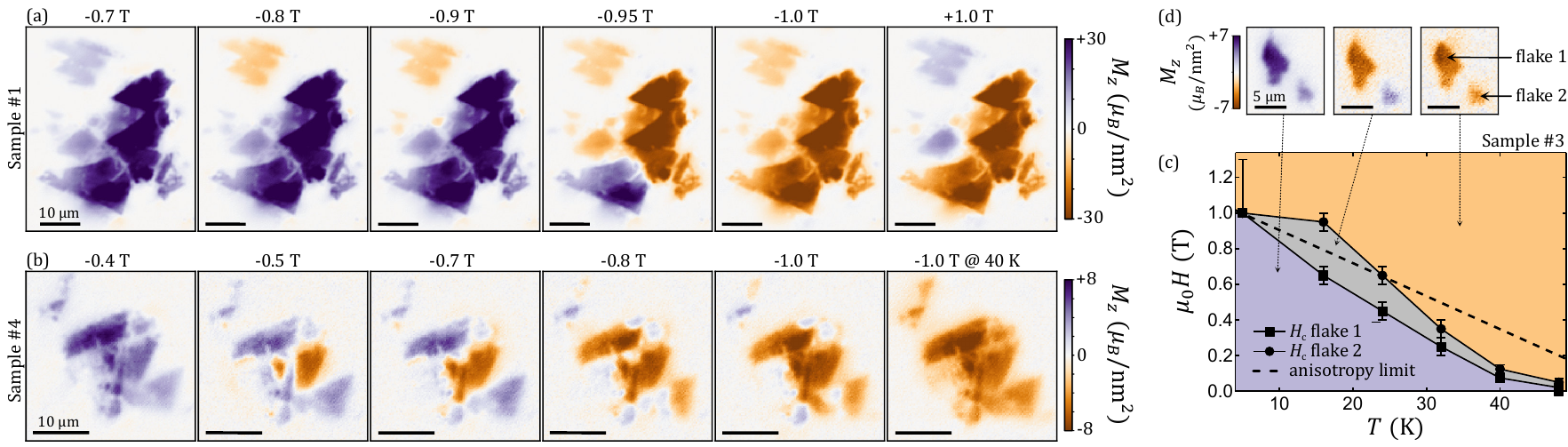}
	\caption{\textbf{Imaging domain reversal in few-layer VI$_3$ flakes.} (a) $M_z$ maps of sample \#1 for increasing magnetic field pulse amplitude from $-0.7$~T to $-1.0$~T (from left to right) applied along the $-z$ direction (pulse duration $\sim10$~s excluding rise/fall times), starting with the flakes magnetized in the $+z$ direction. In the last image of the series, a $+1.0$~T pulse was applied to reverse the magnetization back to its original sign. (b) $M_z$ maps of sample \#4 after pulse amplitudes from $-0.4$~T to $-1.0$~T. In the last image of the series, a $-1.0$~T pulse was applied while heating the sample to $T=40$~K to facilitate magnetization reversal. (c) $H-T$ phase diagram of the magnetic state of two flakes in sample \#3 constructed from image series similar to those in (a,b) at various temperatures (see Fig. S11). The data points indicate the coercive field $H_c$ for each flake. The error bars correspond to the step size in the field amplitude. The large error bar on the lowest temperature points denotes the fact only a lower bound for $H_c$ is determined in this case. The dashed line is the Stoner-Wohlfarth model for the coercive field~\cite{Coey2010} using the temperature-dependent anisotropy constant measured for bulk VI$_3$ in Ref.~\cite{Yan2019}. The purple (orange) shaded region corresponds to the two flakes magnetized in the $+z$ ($-z$) direction, while in the grey region only the largest flake has switched. (d) Example $M_z$ maps corresponding to each magnetic state, recorded at 5 K.
	}  
	\label{Fig3}
\end{figure*}

\begin{figure*}[t!]
	\includegraphics[width=1\textwidth]{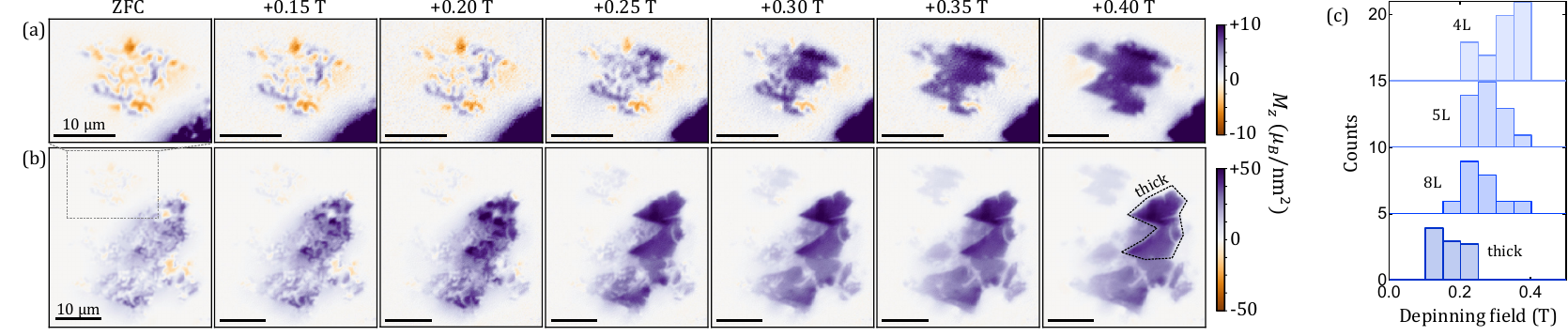}
	\caption{\textbf{Imaging the initial magnetization of few-layer VI$_3$ flakes.} (a,b) $M_z$ maps of sample \#1 as a function of the magnetic field pulse amplitude starting from the virgin state (zero-field cooling, ZFC). (a) is a magnified version of (b) highlighting the region containing flakes from 4 to 8 layer thick. (c) Histograms of the number of domain wall jumps versus magnetic field for flakes of different thickness, constructed from (a) for the thin flakes (4L, 5L, 8L) and from (b) for thick flakes (10-20 nm thickness, dashed box).       
	}  
	\label{Fig4}
\end{figure*}

In hard magnetic materials, the switching process is governed by either the nucleation or pinning of domain walls~\cite{Coey2010}. In bulk magnets, these two mechanisms are normally distinguished by their initial magnetization curves. The domain walls move freely in a nucleation-type magnet, which has a high initial susceptibility, while they are constantly being trapped in a pinning-type magnet, so the initial susceptibility is small until the depinning field is reached. To determine the limiting mechanism in ultrathin VI$_3$, we applied pulses ($\sim10$~s duration) of magnetic field in the $-z$ direction to samples initially magnetized in the $+z$ direction, and imaged the magnetization in a low field after each pulse. Series of images after pulses of increasing amplitude up to $-1$~T are shown in Figs. \ref{Fig3}a and \ref{Fig3}b for samples \#1 and \#4, respectively (see additional data in Fig. S12). The magnetization is observed to reverse abruptly, that is, regions of contiguous material switch sign completely at once rather than creating partially reversed domains. This is the signature of a nucleation-type magnet. The switching field, which corresponds to the coercive field $H_c$ in this case, lies in the range $0.5-1$~T for most flakes, with no apparent correlation with flake thickness. The bilayer flake in sample \#4, for instance, lies in the middle of this range, with a coercive field of $0.7-0.8$~T. Some flakes have an even larger coercive field, as evidenced by the purple domains remaining in sample \#4 after applying $-1.0$~T (which is the maximum field amplitude we can apply in our setup), or by the orange domains in sample \#1 after applying $+1.0$~T. 

By heating the sample to 40 K during the application of the pulse, we were able to switch all of the flakes as shown in Fig. \ref{Fig3}b (last image of the series), suggesting that the coercive field decreases with increasing temperature. Repeating the magnetic field sweep at different temperatures (see full image series in Fig. S11) allows us to form a phase diagram from which the $H_c-T$ relation can be identified. This is shown in Fig.~\ref{Fig3}c for two flakes in sample \#3, with example $M_z$ maps in the different states shown in Fig.~\ref{Fig3}d. For both flakes, $H_c$ decreases monotonically from 1~T to zero when $T$ is increased from 5~K to 50~K. The values are in broad agreement with the Stoner-Wohlfarth model~\cite{Coey2010} (dashed line in Fig.~\ref{Fig3}c) in which the coercive field is only limited by the strength of the perpendicular anisotropy (see details in SI, section IX). These results indicate that the coercivity of ultrathin VI$_3$ is mostly governed by anisotropy-limited domain wall nucleation processes, and hence that magnetization reversal occurs in a near-coherent regime.       

To confirm this picture, we directly image the initial magnetization starting from the virgin state, as shown in Figs. \ref{Fig4}a and \ref{Fig4}b for sample \#1 (see additional data in Fig. S13). The multi-domain structure visible after zero-field cooling progressively disappears when magnetic field pulses of increasing amplitude are applied. Histograms of the domain wall depinning field constructed from these images (Fig.~\ref{Fig4}c) indicate depinning fields in the range $0.1-0.4$~T (at 5 K), with a decreasing trend with increasing thickness. These values together with the thickness dependence are consistent with cracks extending through a single (or a few) atomic layer (see SI, section X). Remarkably, however, these defects do not affect the switching process since the corresponding depinning fields are much smaller than $H_c$, set by the nucleation field.

In summary, we employed widefield NV microscopy to directly reveal the mechanism governing magnetic switching in ultrathin VI$_3$, a van der Waals magnetic semiconductor. Our images of domain reversal indicate that ultrathin VI$_3$, down to two atomic layers, is a nucleation-type hard ferromagnet. This is confirmed by images of the initial magnetization revealing domain wall depinning fields far below the switching field. These experiments establish widefield NV microscopy as a powerful tool for magnetic imaging that can be applied to virtually any van der Waals material or heterostructure. By allowing rapid, quantitative imaging of many samples in parallel, it may facilitate the discovery of novel 2D magnetic materials~\cite{Mounet2018,Gong2019} and the investigation of magnetic processes including skyrmionics, current-driven domain wall motion and other spintronic phenomena~\cite{Fert2013,Emori2013,Lin2019,Shi2019,Han2019,Ding2020}.

\section*{Acknowledgements}

The authors thank Marcus Doherty and Yuerui Lu for stimulating discussions. The work by the University of Melbourne team was supported by the Australian Research Council (ARC) through grants DE170100129, CE170100012, LE180100037 and DP190101506. B.C.J. acknowledges the AFAiiR node of the NCRIS Heavy Ion Capability for access to ion-implantation facilities. D.A.B. and S.E.L. are supported by an Australian Government Research Training Program Scholarship. The work by the RMIT team was supported by the ARC through grant CE170100039, and was performed in part at the RMIT Micro Nano Research Facility (MNRF) in the Victorian Node of the Australian National Fabrication Facility (ANFF) and the RMIT Microscopy and Microanalysis Facility (RMMF). S.T., C.L. and H.L. are supported by the National Key R\&D Program of China (Grants No. 2018YFE0202600, 2016YFA0300504) and the National Natural Science Foundation of China (Grant No. 11774423, 11822412). Z.W. thanks the computing resources of Tianhe II and the Arcuda super-computer in Skoltech, and acknowledges support by the National Natural Science Foundation of China (Grant No. 11604159). 

\bibliographystyle{naturemag}
\bibliography{bib}

\begin{thebibliography}{10}
\expandafter\ifx\csname url\endcsname\relax
  \def\url#1{\texttt{#1}}\fi
\expandafter\ifx\csname urlprefix\endcsname\relax\def\urlprefix{URL }\fi
\providecommand{\bibinfo}[2]{#2}
\providecommand{\eprint}[2][]{\url{#2}}

\bibitem{Huang2017}
\bibinfo{author}{Huang, B.} \emph{et~al.}
\newblock \bibinfo{title}{{Layer-dependent ferromagnetism in a van der Waals
  crystal down to the monolayer limit}}.
\newblock \emph{\bibinfo{journal}{Nature}} \textbf{\bibinfo{volume}{546}},
  \bibinfo{pages}{270--273} (\bibinfo{year}{2017}).

\bibitem{Gong2017}
\bibinfo{author}{Gong, C.} \emph{et~al.}
\newblock \bibinfo{title}{{Discovery of intrinsic ferromagnetism in
  two-dimensional van der Waals crystals}}.
\newblock \emph{\bibinfo{journal}{Nature}} \textbf{\bibinfo{volume}{546}},
  \bibinfo{pages}{265--269} (\bibinfo{year}{2017}).

\bibitem{Burch2018}
\bibinfo{author}{Burch, K.}, \bibinfo{author}{Mandrus, D.} \&
  \bibinfo{author}{Park, J.-G.}
\newblock \bibinfo{title}{{Magnetism in two-dimensional van der Waals
  materials}}.
\newblock \emph{\bibinfo{journal}{Nature}} \textbf{\bibinfo{volume}{563}},
  \bibinfo{pages}{47--52} (\bibinfo{year}{2018}).

\bibitem{Gong2019}
\bibinfo{author}{Gong, C.} \& \bibinfo{author}{Zhang, X.}
\newblock \bibinfo{title}{Two-dimensional magnetic crystals and emergent
  heterostructure devices}.
\newblock \emph{\bibinfo{journal}{Science}} \textbf{\bibinfo{volume}{363}},
  \bibinfo{pages}{eaav4450} (\bibinfo{year}{2019}).

\bibitem{Gibertini2019}
\bibinfo{author}{Gibertini, M.}, \bibinfo{author}{Koperski, M.},
  \bibinfo{author}{Morpurgo, A.} \& \bibinfo{author}{Novoselov, K.}
\newblock \bibinfo{title}{{Magnetic 2D materials and heterostructures}}.
\newblock \emph{\bibinfo{journal}{Nature Nanotechnology}}
  \textbf{\bibinfo{volume}{14}}, \bibinfo{pages}{408--419}
  (\bibinfo{year}{2019}).

\bibitem{Jiang2018}
\bibinfo{author}{Jiang, S.}, \bibinfo{author}{Li, L.}, \bibinfo{author}{Wang,
  Z.}, \bibinfo{author}{Mak, K.~F.} \& \bibinfo{author}{Shan, J.}
\newblock \bibinfo{title}{{Controlling magnetism in 2D CrI$_3$ by electrostatic
  doping}}.
\newblock \emph{\bibinfo{journal}{Nature Nanotechnology}}
  \textbf{\bibinfo{volume}{13}}, \bibinfo{pages}{549} (\bibinfo{year}{2018}).

\bibitem{Huang2018}
\bibinfo{author}{Huang, B.} \emph{et~al.}
\newblock \bibinfo{title}{{Electrical control of 2D magnetism in bilayer
  CrI$_3$}}.
\newblock \emph{\bibinfo{journal}{Nature Nanotechnology}}
  \textbf{\bibinfo{volume}{13}}, \bibinfo{pages}{544--548}
  (\bibinfo{year}{2018}).

\bibitem{Li2019}
\bibinfo{author}{Li, T.} \emph{et~al.}
\newblock \bibinfo{title}{{Pressure-controlled interlayer magnetism in
  atomically thin CrI$_3$}}.
\newblock \emph{\bibinfo{journal}{Nature Materials}}
  \textbf{\bibinfo{volume}{18}}, \bibinfo{pages}{1303} (\bibinfo{year}{2019}).

\bibitem{Chen2019}
\bibinfo{author}{Chen, W.} \emph{et~al.}
\newblock \bibinfo{title}{{Direct observation of van der Waals
  stacking{\textendash}dependent interlayer magnetism}}.
\newblock \emph{\bibinfo{journal}{Science}} \textbf{\bibinfo{volume}{366}},
  \bibinfo{pages}{983--987} (\bibinfo{year}{2019}).

\bibitem{McGuire2015}
\bibinfo{author}{McGuire, M.~A.}, \bibinfo{author}{Dixit, H.},
  \bibinfo{author}{Cooper, V.~R.} \& \bibinfo{author}{Sales, B.~C.}
\newblock \bibinfo{title}{Coupling of crystal structure and magnetism in the
  layered, ferromagnetic insulator {CrI$_3$}}.
\newblock \emph{\bibinfo{journal}{Chemistry of Materials}}
  \textbf{\bibinfo{volume}{27}}, \bibinfo{pages}{612--620}
  (\bibinfo{year}{2015}).

\bibitem{Chen2013}
\bibinfo{author}{Chen, B.} \emph{et~al.}
\newblock \bibinfo{title}{Magnetic properties of layered itinerant electron
  ferromagnet {Fe$_3$GeTe$_2$}}.
\newblock \emph{\bibinfo{journal}{Journal of the Physical Society of Japan}}
  \textbf{\bibinfo{volume}{82}}, \bibinfo{pages}{124711}
  (\bibinfo{year}{2013}).

\bibitem{Tan2018}
\bibinfo{author}{Tan, C.} \emph{et~al.}
\newblock \bibinfo{title}{{Hard magnetic properties in nanoflake van der Waals
  Fe$_3$GeTe$_2$}}.
\newblock \emph{\bibinfo{journal}{Nature Communications}}
  \textbf{\bibinfo{volume}{9}}, \bibinfo{pages}{1554} (\bibinfo{year}{2018}).

\bibitem{Fei2018}
\bibinfo{author}{Fei, Z.} \emph{et~al.}
\newblock \bibinfo{title}{Two-dimensional itinerant ferromagnetism in
  atomically thin {Fe$_3$GeTe$_2$}}.
\newblock \emph{\bibinfo{journal}{Nature Materials}}
  \textbf{\bibinfo{volume}{17}}, \bibinfo{pages}{778--782}
  (\bibinfo{year}{2018}).

\bibitem{Deng2018}
\bibinfo{author}{Deng, Y.} \emph{et~al.}
\newblock \bibinfo{title}{Gate-tunable room-temperature ferromagnetism in
  two-dimensional {Fe$_3$GeTe$_2$}}.
\newblock \emph{\bibinfo{journal}{Nature}} \textbf{\bibinfo{volume}{563}},
  \bibinfo{pages}{94} (\bibinfo{year}{2018}).

\bibitem{LeSage2013}
\bibinfo{author}{{Le Sage}, D.} \emph{et~al.}
\newblock \bibinfo{title}{{Optical magnetic imaging of living cells}}.
\newblock \emph{\bibinfo{journal}{Nature}} \textbf{\bibinfo{volume}{496}},
  \bibinfo{pages}{486--9} (\bibinfo{year}{2013}).

\bibitem{Tetienne2017}
\bibinfo{author}{Tetienne, J.-P.} \emph{et~al.}
\newblock \bibinfo{title}{{Quantum imaging of current flow in graphene}}.
\newblock \emph{\bibinfo{journal}{Sci. Adv.}} \textbf{\bibinfo{volume}{3}},
  \bibinfo{pages}{e1602429} (\bibinfo{year}{2017}).

\bibitem{Casola2018}
\bibinfo{author}{Casola, F.}, \bibinfo{author}{{Van Der Sar}, T.} \&
  \bibinfo{author}{Yacoby, A.}
\newblock \bibinfo{title}{{Probing condensed matter physics with magnetometry
  based on nitrogen-vacancy centres in diamond}}.
\newblock \emph{\bibinfo{journal}{Nature Reviews Materials}}
  \textbf{\bibinfo{volume}{3}}, \bibinfo{pages}{17088} (\bibinfo{year}{2018}).

\bibitem{Lillie2020}
\bibinfo{author}{Lillie, S.~E.} \emph{et~al.}
\newblock \bibinfo{title}{Laser modulation of superconductivity in a cryogenic
  widefield nitrogen-vacancy microscope}.
\newblock \emph{\bibinfo{journal}{Nano Letters}} \textbf{\bibinfo{volume}{20}},
  \bibinfo{pages}{1855--1861} (\bibinfo{year}{2020}).

\bibitem{Tian2019}
\bibinfo{author}{Tian, S.} \emph{et~al.}
\newblock \bibinfo{title}{{Ferromagnetic van der Waals Crystal VI$_3$}}.
\newblock \emph{\bibinfo{journal}{Journal of the American Chemical Society}}
  \textbf{\bibinfo{volume}{141}}, \bibinfo{pages}{5326--5333}
  (\bibinfo{year}{2019}).

\bibitem{Son2019}
\bibinfo{author}{Son, S.} \emph{et~al.}
\newblock \bibinfo{title}{{Bulk properties of the van der Waals hard
  ferromagnet ${\mathrm{VI}}_{3}$}}.
\newblock \emph{\bibinfo{journal}{Phys. Rev. B}} \textbf{\bibinfo{volume}{99}},
  \bibinfo{pages}{041402} (\bibinfo{year}{2019}).

\bibitem{Kong2019}
\bibinfo{author}{Kong, T.} \emph{et~al.}
\newblock \bibinfo{title}{{VI$_3$ -- a New Layered Ferromagnetic
  Semiconductor}}.
\newblock \emph{\bibinfo{journal}{Advanced Materials}}
  \textbf{\bibinfo{volume}{31}}, \bibinfo{pages}{1808074}
  (\bibinfo{year}{2019}).

\bibitem{Thiel2019}
\bibinfo{author}{Thiel, L.} \emph{et~al.}
\newblock \bibinfo{title}{{Probing magnetism in 2D materials at the nanoscale
  with single-spin microscopy}}.
\newblock \emph{\bibinfo{journal}{Science}} \textbf{\bibinfo{volume}{364}},
  \bibinfo{pages}{973–976} (\bibinfo{year}{2019}).

\bibitem{Coey2010}
\bibinfo{author}{Coey, J. M.~D.}
\newblock \emph{\bibinfo{title}{{Magnetism and Magnetic Materials}}}
  (\bibinfo{publisher}{Cambridge University Press}, \bibinfo{year}{2010}).

\bibitem{Yan2019}
\bibinfo{author}{Yan, J.} \emph{et~al.}
\newblock \bibinfo{title}{{Anisotropic magnetic entropy change in the hard
  ferromagnetic semiconductor $\mathrm{V}{\mathrm{I}}_{3}$}}.
\newblock \emph{\bibinfo{journal}{Phys. Rev. B}}
  \textbf{\bibinfo{volume}{100}}, \bibinfo{pages}{094402}
  (\bibinfo{year}{2019}).

\bibitem{Mounet2018}
\bibinfo{author}{Mounet, N.} \emph{et~al.}
\newblock \bibinfo{title}{{Two-dimensional materials from high-throughput
  computational exfoliation of experimentally known compounds}}.
\newblock \emph{\bibinfo{journal}{Nature Nanotechnology}}
  \textbf{\bibinfo{volume}{13}}, \bibinfo{pages}{246--252}
  (\bibinfo{year}{2018}).

\bibitem{Fert2013}
\bibinfo{author}{Fert, A.}, \bibinfo{author}{Cros, V.} \&
  \bibinfo{author}{Sampaio, J.}
\newblock \bibinfo{title}{{Skyrmions on the track}}.
\newblock \emph{\bibinfo{journal}{Nature Nanotechnology}}
  \textbf{\bibinfo{volume}{8}}, \bibinfo{pages}{152--156}
  (\bibinfo{year}{2013}).

\bibitem{Emori2013}
\bibinfo{author}{Emori, S.}, \bibinfo{author}{Bauer, U.}, \bibinfo{author}{Ahn,
  S.-M.}, \bibinfo{author}{Martinez, E.} \& \bibinfo{author}{Beach, G.}
\newblock \bibinfo{title}{{Current-driven dynamics of chiral ferromagnetic
  domain walls}}.
\newblock \emph{\bibinfo{journal}{Nature Materials}}
  \textbf{\bibinfo{volume}{12}}, \bibinfo{pages}{611} (\bibinfo{year}{2013}).

\bibitem{Lin2019}
\bibinfo{author}{Xiaoyang~Lin, K. L.~W., Wei~Yang} \& \bibinfo{author}{Zhao,
  W.}
\newblock \bibinfo{title}{Two-dimensional spintronics for low-power
  electronics}.
\newblock \emph{\bibinfo{journal}{Nature Electronics}}
  \textbf{\bibinfo{volume}{2}}, \bibinfo{pages}{274--283}
  (\bibinfo{year}{2019}).

\bibitem{Shi2019}
\bibinfo{author}{Shi, S.} \emph{et~al.}
\newblock \bibinfo{title}{{All-electric magnetization switching and
  Dzyaloshinskii-Moriya interaction in WTe$_2$/ferromagnet heterostructures}}.
\newblock \emph{\bibinfo{journal}{Nature Nanotechnology}}
  \textbf{\bibinfo{volume}{14}}, \bibinfo{pages}{945--949}
  (\bibinfo{year}{2019}).

\bibitem{Han2019}
\bibinfo{author}{Han, M.-G.} \emph{et~al.}
\newblock \bibinfo{title}{{Topological Magnetic-Spin Textures in
  Two-Dimensional van der Waals Cr2Ge2Te6}}.
\newblock \emph{\bibinfo{journal}{Nano Letters}} \textbf{\bibinfo{volume}{19}},
  \bibinfo{pages}{7859--7865} (\bibinfo{year}{2019}).

\bibitem{Ding2020}
\bibinfo{author}{Ding, B.} \emph{et~al.}
\newblock \bibinfo{title}{{Observation of Magnetic Skyrmion Bubbles in a van
  der Waals Ferromagnet Fe3GeTe2}}.
\newblock \emph{\bibinfo{journal}{Nano Letters}} \textbf{\bibinfo{volume}{20}},
  \bibinfo{pages}{868--873} (\bibinfo{year}{2020}).

\bibitem{Tetienne2018}
\bibinfo{author}{Tetienne, J.-P.} \emph{et~al.}
\newblock \bibinfo{title}{{Spin coherence of dense near-surface ensembles of
  nitrogen-vacancy centres in diamond}}.
\newblock \emph{\bibinfo{journal}{Phys. Rev. B}} \textbf{\bibinfo{volume}{97}},
  \bibinfo{pages}{085402} (\bibinfo{year}{2018}).

\bibitem{Zomer2014}
\bibinfo{author}{Zomer, P.~J.}, \bibinfo{author}{Guimarães, M. H.~D.},
  \bibinfo{author}{Brant, J.~C.}, \bibinfo{author}{Tombros, N.} \&
  \bibinfo{author}{van Wees, B.~J.}
\newblock \bibinfo{title}{Fast pick up technique for high quality
  heterostructures of bilayer graphene and hexagonal boron nitride}.
\newblock \emph{\bibinfo{journal}{Applied Physics Letters}}
  \textbf{\bibinfo{volume}{105}}, \bibinfo{pages}{013101}
  (\bibinfo{year}{2014}).

\bibitem{Doherty2013}
\bibinfo{author}{Doherty, M.~W.} \emph{et~al.}
\newblock \bibinfo{title}{{The nitrogen-vacancy colour centre in diamond}}.
\newblock \emph{\bibinfo{journal}{Physics Reports}}
  \textbf{\bibinfo{volume}{528}}, \bibinfo{pages}{1--45}
  (\bibinfo{year}{2013}).

\bibitem{Rondin2014}
\bibinfo{author}{Rondin, L.} \emph{et~al.}
\newblock \bibinfo{title}{{Magnetometry with nitrogen-vacancy defects in
  diamond}}.
\newblock \emph{\bibinfo{journal}{Rep. Prog. Phys.}}
  \textbf{\bibinfo{volume}{77}}, \bibinfo{pages}{56503} (\bibinfo{year}{2014}).

\bibitem{Rondin2012}
\bibinfo{author}{Rondin, L.} \emph{et~al.}
\newblock \bibinfo{title}{Nanoscale magnetic field mapping with a single spin
  scanning probe magnetometer}.
\newblock \emph{\bibinfo{journal}{Applied Physics Letters}}
  \textbf{\bibinfo{volume}{100}}, \bibinfo{pages}{153118}
  (\bibinfo{year}{2012}).

\bibitem{Glass2006}
\bibinfo{author}{Glass, C.~W.}, \bibinfo{author}{Oganov, A.~R.} \&
  \bibinfo{author}{Hansen, N.}
\newblock \bibinfo{title}{{USPEX-Evolutionary crystal structure prediction}}.
\newblock \emph{\bibinfo{journal}{Computer Physics Communications}}
  \textbf{\bibinfo{volume}{175}}, \bibinfo{pages}{713 -- 720}
  (\bibinfo{year}{2006}).

\bibitem{Oganov2006}
\bibinfo{author}{Oganov, A.~R.} \& \bibinfo{author}{Glass, C.~W.}
\newblock \bibinfo{title}{Crystal structure prediction using ab initio
  evolutionary techniques: Principles and applications}.
\newblock \emph{\bibinfo{journal}{The Journal of Chemical Physics}}
  \textbf{\bibinfo{volume}{124}}, \bibinfo{pages}{244704}
  (\bibinfo{year}{2006}).

\bibitem{Kresse1993}
\bibinfo{author}{Kresse, G.} \& \bibinfo{author}{Hafner, J.}
\newblock \bibinfo{title}{Ab initio molecular dynamics for open-shell
  transition metals}.
\newblock \emph{\bibinfo{journal}{Phys. Rev. B}} \textbf{\bibinfo{volume}{48}},
  \bibinfo{pages}{13115--13118} (\bibinfo{year}{1993}).

\bibitem{Kresse1996}
\bibinfo{author}{Kresse, G.} \& \bibinfo{author}{Furthmüller, J.}
\newblock \bibinfo{title}{Efficiency of ab-initio total energy calculations for
  metals and semiconductors using a plane-wave basis set}.
\newblock \emph{\bibinfo{journal}{Computational Materials Science}}
  \textbf{\bibinfo{volume}{6}}, \bibinfo{pages}{15 -- 50}
  (\bibinfo{year}{1996}).

\bibitem{Kresse1999}
\bibinfo{author}{Kresse, G.} \& \bibinfo{author}{Joubert, D.}
\newblock \bibinfo{title}{From ultrasoft pseudopotentials to the projector
  augmented-wave method}.
\newblock \emph{\bibinfo{journal}{Phys. Rev. B}} \textbf{\bibinfo{volume}{59}},
  \bibinfo{pages}{1758--1775} (\bibinfo{year}{1999}).

\bibitem{Perdew1996}
\bibinfo{author}{Perdew, J.~P.}, \bibinfo{author}{Burke, K.} \&
  \bibinfo{author}{Ernzerhof, M.}
\newblock \bibinfo{title}{Generalized gradient approximation made simple}.
\newblock \emph{\bibinfo{journal}{Phys. Rev. Lett.}}
  \textbf{\bibinfo{volume}{77}}, \bibinfo{pages}{3865--3868}
  (\bibinfo{year}{1996}).

\bibitem{Grimme2006}
\bibinfo{author}{Grimme, S.}
\newblock \bibinfo{title}{{Semiempirical GGA-type density functional
  constructed with a long-range dispersion correction}}.
\newblock \emph{\bibinfo{journal}{Journal of Computational Chemistry}}
  \textbf{\bibinfo{volume}{27}}, \bibinfo{pages}{1787--1799}
  (\bibinfo{year}{2006}).

\bibitem{Ong2013}
\bibinfo{author}{Ong, S.~P.} \emph{et~al.}
\newblock \bibinfo{title}{Python materials genomics (pymatgen): A robust,
  open-source python library for materials analysis}.
\newblock \emph{\bibinfo{journal}{Computational Materials Science}}
  \textbf{\bibinfo{volume}{68}}, \bibinfo{pages}{314 -- 319}
  (\bibinfo{year}{2013}).

\bibitem{He2016}
\bibinfo{author}{He, J.}, \bibinfo{author}{Ma, S.}, \bibinfo{author}{Lyu, P.}
  \& \bibinfo{author}{Nachtigall, P.}
\newblock \bibinfo{title}{Unusual dirac half-metallicity with intrinsic
  ferromagnetism in vanadium trihalide monolayers}.
\newblock \emph{\bibinfo{journal}{J. Mater. Chem. C}}
  \textbf{\bibinfo{volume}{4}}, \bibinfo{pages}{2518--2526}
  (\bibinfo{year}{2016}).

\end{thebibliography}

\clearpage

\renewcommand{\thefigure}{S\arabic{figure}}
\renewcommand{\thetable}{S\arabic{table}}
\setcounter{figure}{0}

\begin{widetext}

\section*{Supplementary Information}

\section{Diamond samples} \label{sec:diamonds}

The NV-diamond substrates used in this work were made from $4.4$\,mm~$\times$~$4.4$\,mm~$\times$~$50$\,$\mu$m type-Ib, single-crystal diamond substrates grown by high-pressure, high-temperature synthesis, with $\{100\}$-oriented polished faces (best surface roughness $<5$\,nm Ra), purchased from Delaware Diamond Knives. The diamonds had an initial nitrogen concentration of the order of [N]~$\sim 100$\,ppm. To create vacancies, the received plates were irradiated with $^{12}$C$^-$ ions accelerated at $100$\,keV with a fluence of $10^{12}-10^{13}$ ions/cm$^2$. We performed full cascade Stopping and Range of Ions in Matter (SRIM) Monte Carlo simulations to estimate the depth distribution of the created vacancies (Fig.~\ref{FigSI_diamonds}a), predicting a distribution spanning the range $0$~-~$200$\,nm with a peak vacancy density of $\sim110$~ppm (for a $10^{12}$ ions/cm$^2$ fluence) at a depth of $\sim 130$\,nm. Following irradiation the diamonds were laser cut into smaller $2.2$\,mm~$\times$~$2.2$\,mm~$\times$~$50$\,$\mu$m plates, which were then annealed in a vacuum of $\sim10^{-5}$~Torr to form the NV centres, using the following sequence~\cite{Tetienne2018}: $6$\,h at $400$\,$^\circ$C, $6$\,h ramp to $800$\,$^\circ$C, $6$\,h at $800$\,$^\circ$C, $6$\,h ramp to $1100$\,$^\circ$C, $2$\,h at $1100$\,$^\circ$C, $2$\,h ramp to room temperature. After annealing the plates were acid cleaned ($15$\,minutes in a boiling mixture of sulphuric acid and sodium nitrate).

\begin{figure}[b!]
	\begin{center}
		\includegraphics[width=0.6\textwidth]{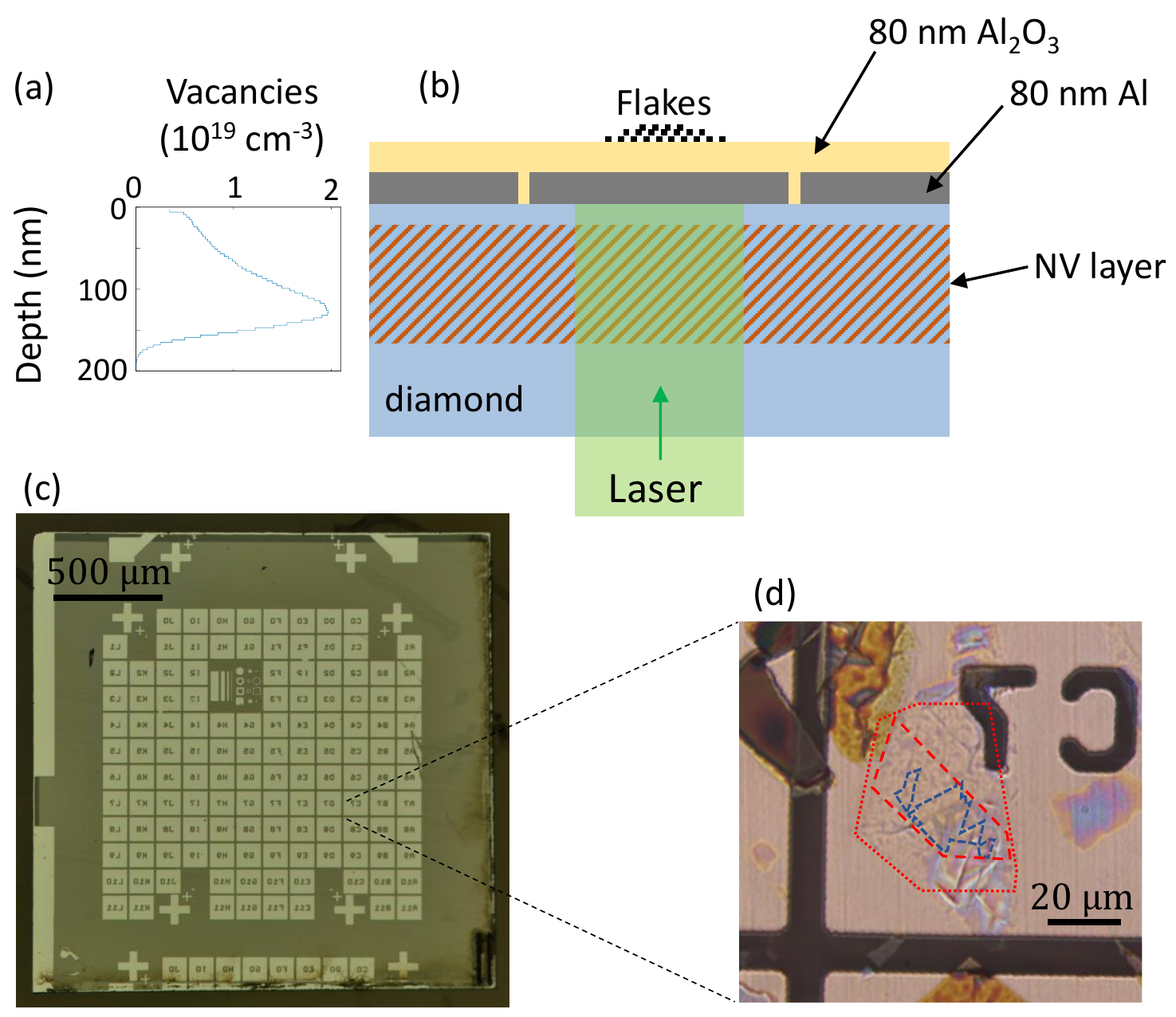}
		\caption{\textbf{NV-diamond substrates.} (a) Vacancy concentration as a function of depth for a $100$\,keV $^{12}$C$^-$ implant in diamond at a dose of $10^{12}$\,ions/cm$^2$, calculated from Stopping and Range of Ions in Matter (SRIM) simulations. We assumed a diamond density of $3.51$\,g\,cm$^{-3}$ and a displacement energy of $50$\,eV. The peak vacancy concentration of $\sim2\times10^{19}$~cm$^{-3}$ corresponds to $\sim110$~ppm. (b) Schematic cross-section of the substrates used for NV imaging. The diamond is coated with an 80 nm patterned Al layer capped with an 80 nm Al$_2$O$_3$ layer. (c) Optical micrograph of a typical diamond substrate with this Al/Al$_2$O$_3$ structure. (d) Optical micrograph of a substrate after transferring VI$_3$ flakes (blue dashed lines) encapsulated with hBN flakes (red dashed/dotted lines).}
		\label{FigSI_diamonds}
	\end{center}
\end{figure}

To facilitate localisation of the VI$_3$ flakes in the NV microscope, metallic grids were fabricated on the diamond substrates by photolithography, thermal evaporation of 80 nm of Al, and lift-off. The substrates were then coated with 80 nm of Al$_2$O$_3$ by atomic layer deposition, providing a relatively smooth surface (surface roughness $<2$\,nm Ra measured by atomic force microscopy) for the subsequent transfer of VI$_3$ flakes (Fig.~\ref{FigSI_diamonds}b-d). This Al/Al$_2$O$_3$ structure serves several additional purposes. First, the Al layer acts as a laser shield during NV imaging, preventing the laser light from reaching the VI$_3$ samples. The majority of the laser light is reflected at the diamond/Al interface, with only a small fraction ($<20\%$) being absorbed in the Al layer, generating heat. The Al$_2$O$_3$ acts as a thermal insulator to minimise heating of the VI$_3$ flakes. Indeed, at 4 K Al and diamond are much better thermal conductors than Al$_2$O$_3$ and hBN along the $c$ axis (the ratio of thermal conductivities is about a hundred) and so the heat generated by the laser in the Al layer is expected to dissipate into the substrate (and eventually into the He exchange gas) rather than towards the VI$_3$ flakes. Finally, the thickness of the Al$_2$O$_3$ layer (80 nm) was chosen to maximise the optical contrast of atomically thin materials deposited on the Al/Al$_2$O$_3$ structure, in a similar fashion to standard Si/SiO$_2$(285 nm) substrates, thus facilitating sample preparation.

\section{VI$_3$ samples}

The VI$_3$ flakes studied in this work were obtained by mechanical exfoliation of a bulk VI$_3$ crystal grown by the chemical vapour transport method. The growth details and bulk magnetic properties were reported previously~\cite{Tian2019}. The entire heterostructure assembly was carried out in a glove box (H$_2$O and O$_2 < 0.1$~ppm) filled with Ar. Firstly, hBN flakes of appropriate thickness (5-15 nm) were prepared and identified on Si/SiO$_2$(285 nm) substrates. Each selected hBN flake was scanned by an atomic force microscope (AFM) to reveal its cleanliness and thickness. Secondly, tested substrates (with hBN on) and cleaned blank Si/SiO2 substrates were baked in the glove box for 5 minutes at 120$^\circ$C to minimize the potential for water adhesion. After baking, the substrates were put aside, pending further heterostructure stacking and mechanical exfoliation. Thirdly, thin VI$_3$ flakes were prepared and optically identified on blank Si/SiO$_2$(285 nm) substrates. Then, a standard pick-up technique~\cite{Zomer2014} was utilized to fabricate the hBN/VI$_3$/hBN heterostructures. After layer-by-layer pick-up by a polycarbonate/polydimethylsiloxane (PC/PDMS) stamp, the heterostructures were released onto diamond substrates. The released PC layer remained on top of the samples throughout the subsequent measurements including NV imaging. The results presented in the main text correspond to four different stacks prepared on two different diamond substrates.

\begin{figure*}[b!]
	\begin{center}
		\includegraphics[width=0.9\textwidth]{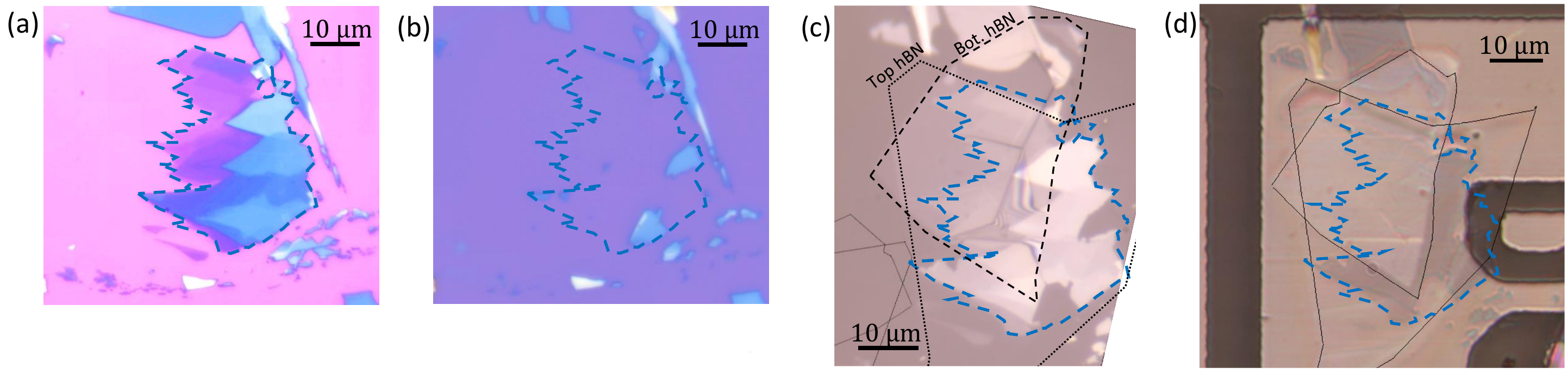}
		\caption{\textbf{Sample \#1.} (a) Optical image of VI$_3$ flakes prepared on a Si/SiO$_2$ substrate. (b) Optical image of the same region of the substrate after picking up the flakes with the PDMS stamp. (c) Optical image of the hBN/VI$_3$/hBN stack on the PDMS stamp. (d) Optical image of the same stack after release onto the diamond substrate. The dashed/dotted lines are guides to the eye representing the VI$_3$ flakes (blue) and the hBN flakes (black).}
		\label{FigSI_sample351}
	\end{center}
\end{figure*}

\begin{figure*}[b!]
	\begin{center}
		\includegraphics[width=0.7\textwidth]{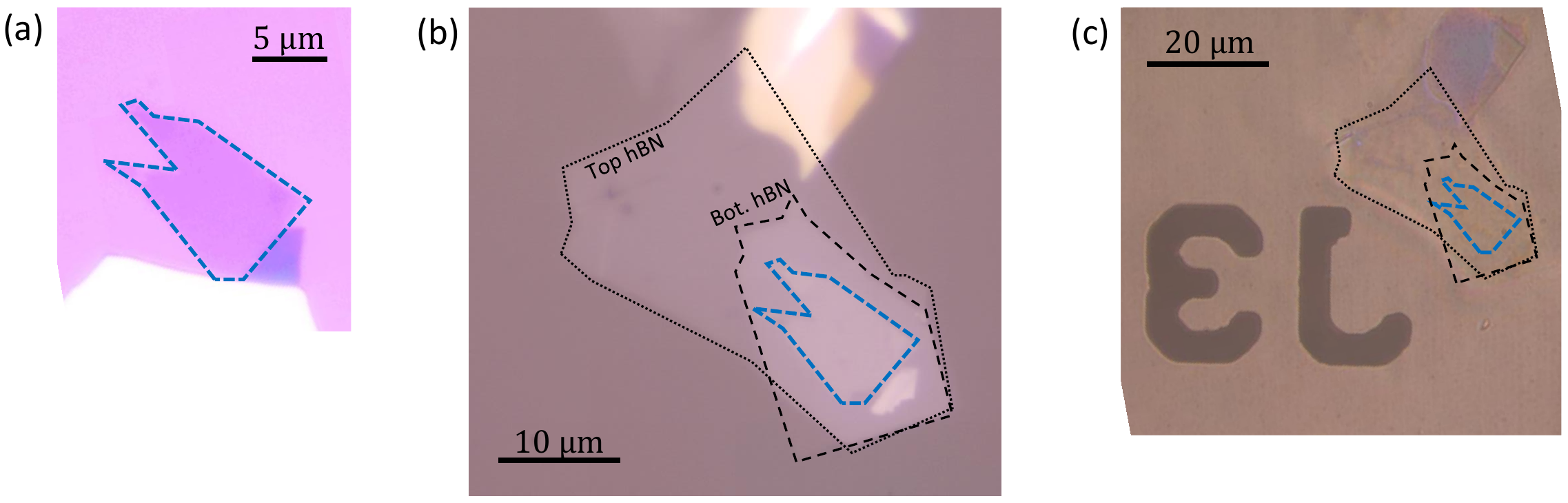}
		\caption{\textbf{Sample \#2.} (a) Optical image of VI$_3$ flakes prepared on a Si/SiO$_2$ substrate. (b) Optical image of the hBN/VI$_3$/hBN stack on the PDMS stamp. (d) Optical image of the same stack after release onto the diamond substrate.}
		\label{FigSI_sample357b1}
	\end{center}
\end{figure*}

\begin{figure*}[t!]
	\begin{center}
		\includegraphics[width=0.8\textwidth]{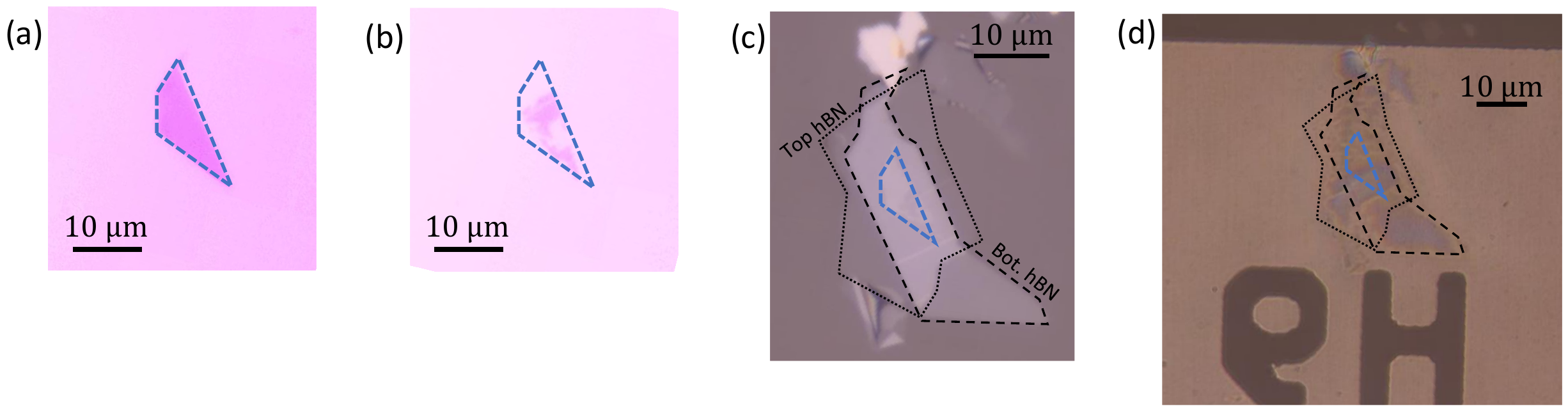}
		\caption{\textbf{Sample \#3.} (a) Optical image of VI$_3$ flakes prepared on a Si/SiO$_2$ substrate. (b) Optical image of the same region of the substrate after picking up the flakes with the PDMS stamp. (c) Optical image of the hBN/VI$_3$/hBN stack on the PDMS stamp. (d) Optical image of the same stack after release onto the diamond substrate.}
		\label{FigSI_sample357b2}
	\end{center}
\end{figure*}

\begin{figure*}[t!]
	\begin{center}
		\includegraphics[width=0.8\textwidth]{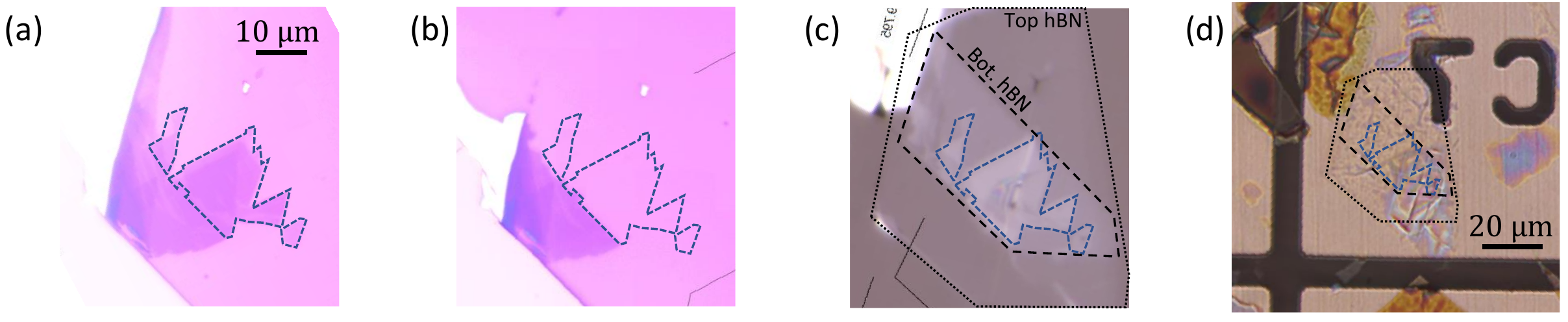}
		\caption{\textbf{Sample \#4.} (a) Optical image of VI$_3$ flakes prepared on a Si/SiO$_2$ substrate. (b) Optical image of the same region of the substrate after picking up the flakes with the PDMS stamp. (c) Optical image of the hBN/VI$_3$/hBN stack on the PDMS stamp. (d) Optical image of the same stack after release onto the diamond substrate.}
		\label{FigSI_sample357b3}
	\end{center}
\end{figure*}

During transfer to the NV microscope, the samples were exposed to air for about 5 minutes, leading to partial oxidation of the uncovered VI$_3$ flakes as evidenced by the fragmentation of some of the magnetic features in Fig. 1b of the main text. Likewise, imperfections in the hBN encapsulation (e.g. cracks or gaps) could lead to localised degradation of the VI$_3$ flakes during this time. This may be the reason for the absence of magnetic signal observed, for instance, in the region comprising the trilayer in sample \#1 (see Fig. 2 of the main text). After loading, the sample chamber of the NV microscope was pumped to a vacuum of $<10^{-4}$~mbar and then filled with a He exchange gas (room temperature pressure of 10-100 mbar). No sign of further sample degradation was observed during the several weeks of NV measurements. 

Optical images were taken at various stages of the stacking process to monitor the integrity of the VI$_3$ flakes, shown in Figs.~\ref{FigSI_sample351}-\ref{FigSI_sample357b3} for samples \#1-4, respectively. Images of the Si/SiO$_2$ substrate after picking up the flakes with the PC/PDMS stamp allow us to determine whether damage or further exfoliation of the picked-up flakes may have occurred. For example, we see that all the thin flakes from sample \#1 were successfully picked up (compare Fig.~\ref{FigSI_sample351}a and Fig.~\ref{FigSI_sample351}b), whereas in sample \#3 some of the layers partially remained on the Si/SiO$_2$ substrate (Fig.~\ref{FigSI_sample357b2}b), which explains the gap in the $M_z$ map in Fig. 2b of the main text. Furthermore, images of the PC/PDMS stamp allow us to trace the location of the hBN flakes relative to the VI$_3$ flakes (see, e.g., Fig.~\ref{FigSI_sample357b1}b). Finally, images of the stacks on the diamond substrate allow spatial correlation with the NV images (see, e.g., Fig.~\ref{FigSI_sample357b1}c).

\section{Thickness of the VI$_3$ flakes}
 
The thickness of the studied VI$_3$ flakes was determined using the relative optical contrast between the VI$_3$ flakes and the Si/SiO$_2$(285 nm) substrate, using the red channel of the images obtained under white light illumination (the red channel gives the highest contrast). The optical contrast was defined as $C=\frac{I_{\rm flake}-I_{\rm substrate}}{I_{\rm flake}+I_{\rm substrate}}$, where $I_{\rm flake}$ ($I_{\rm substrate}$) is the intensity on the flake (substrate)~\cite{Huang2017,Thiel2019}. Example optical contrast maps are shown in Fig.~\ref{FigSI_thickness}a, which includes the four samples studied in the main text. Using an appropriate colour scheme, discrete steps are clearly visible. By analysing a large number of such maps and reading the value $C$ for each sufficiently large domain observable, we identified the contrast from a monolayer of VI$_3$ to be $C_{\rm mono}\approx2.0\%$ in our setup (Fig.~\ref{FigSI_thickness}b). Using this value, the number of layers for a given flake can be estimated as $N={\rm round}(C/C_{\rm mono})$, which is approximately valid for flakes under $\sim10$~nm thickness~\cite{Huang2017}. This number $N$ is indicated on selected flakes in Fig.~\ref{FigSI_thickness}a and was used in the main text to define the thickness of each flake analysed. The optical contrast vs thickness relationship was confirmed by performing atomic force microscopy (AFM) on several VI$_3$ flakes covered with hBN. An example AFM image is shown in Fig.~\ref{FigSI_thickness}c of a 5 to 9 layer flake (based on the optical contrast image shown on the right). The line cut across an edge reveals a 3-4 nm step, in agreement with the step expected for 5 layers of VI$_3$ given the interlayer distance of 0.70 nm at room temperature~\cite{Tian2019}. 

\begin{figure*}[t!]
	\begin{center}
		\includegraphics[width=0.8\textwidth]{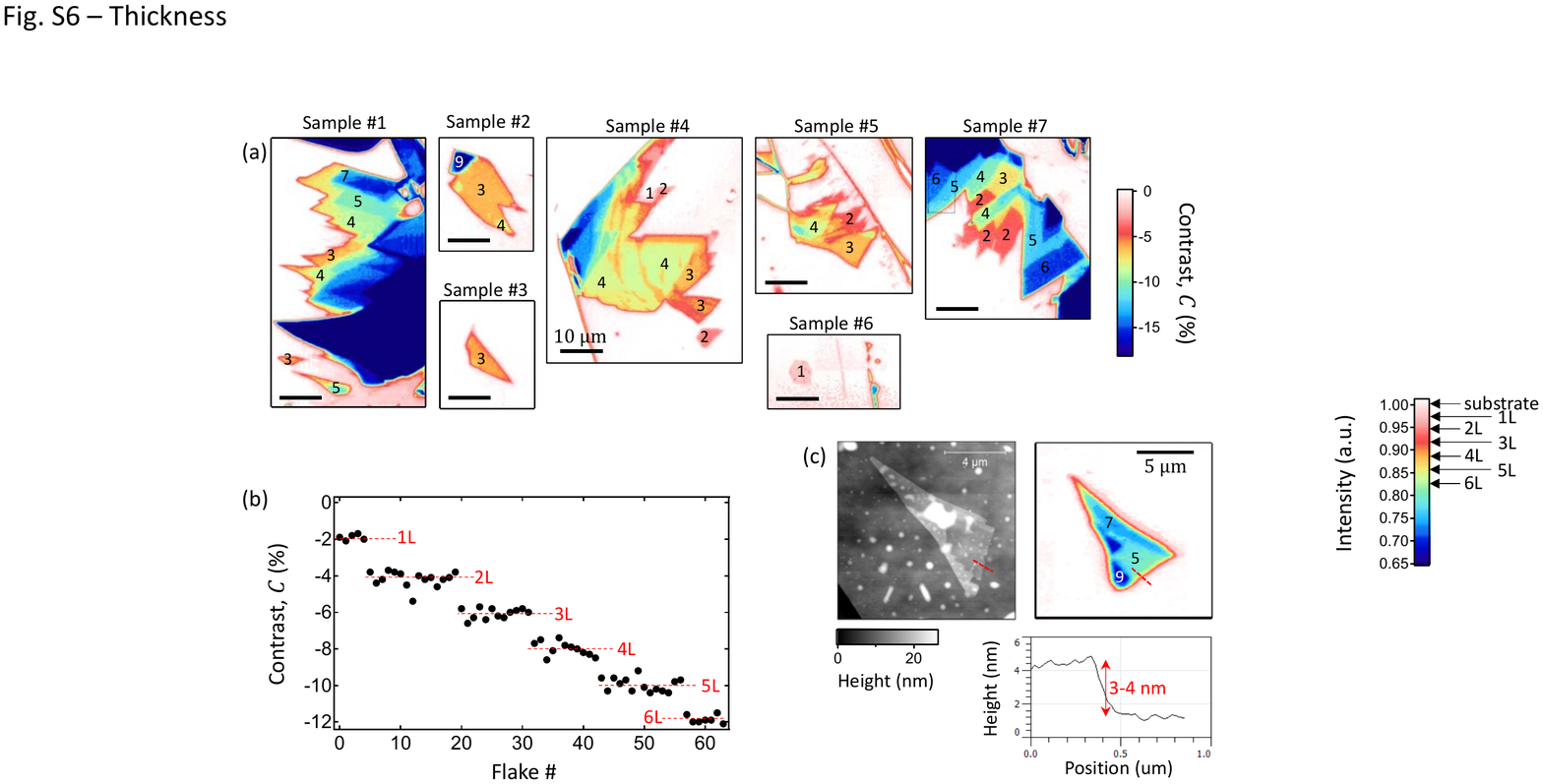}
		\caption{\textbf{Determining the thickness of the VI$_3$ flakes.} (a) Optical contrast maps of VI$_3$ flakes on a Si/SiO$_2$(285 nm) substrate, as defined in the text. The overlaid numbers indicate the estimated number of layers based on the analysis in (b). (b) Contrast of a large number of flakes revealing discrete steps at $-2\%$ (attributed to a single layer of VI$_3$, 1L), at $-4\%$ (two layers, 2L), etc. (c) Atomic force microscope image (left) and optical contrast map (right) of a 5-9 layer flake. The graph shows a line cut across the 5-layer edge.}
		\label{FigSI_thickness}
	\end{center}
\end{figure*}  

\section{NV microscope setup} \label{sec:setup}

The widefield NV microscope used in this work was described previously~\cite{Lillie2020} and is built around a closed-cycle cryostat (Attocube attoDRY1000) integrating a 1-T superconducting vector magnet (Cryomagnetics). Optical excitation was achieved by using a 532~nm continuous wave (CW) laser (Laser Quantum Ventus) coupled to a single-mode fiber and gated with a fibre-coupled acousto-optic modulator (AAOpto MQ180-G9-Fio). The collimation of the laser beam was adjusted in order to produce a $\sim50~\mu$m-wide spot at the NV layer after passing the low-temperature microscope objective (Attocube LT-APO/VISIR/0.82). The laser power entering the cryostat was between 10 and 50 mW for the images shown in the main text, corresponding to a peak intensity of about 20-100 W/cm$^2$ at the NV layer taking into account reflections at the multiple windows and interfaces. The NV photoluminescence (PL) was collected by the same objective, separated from the excitation laser by a dichroic beam splitter, filtered through a $731/137$\,nm band pass filter, and imaged onto a water cooled sCMOS camera (Andor Zyla 5.5-W USB3). 

To allow driving of the NV spin state, the diamond was glued to a glass cover slip patterned with a microwave resonator connected to a printed circuit board (PCB) mounted on a stack of positioners (Attocube ANPxyz101). The microwave signal passed to the PCB was delivered by a signal generator (Rohde \& Schwarz SMB100A), gated by a switch (Mini-Circuits ZASWA-2-50DR+) and amplified (Mini-Circuits HPA-50W-63). All measurements were sequenced using a pulse pattern generator (SpinCore PulseBlasterESR-PRO 500 MHz) to gate the laser and microwave, and synchronize the image acquisition. The spatial resolution of the microscope was estimated to be $\approx700$~nm~\cite{Lillie2020}, limited by optical aberrations due to imaging through the $150$\,$\mu$m thick cover slip and the $50~\mu$m thick diamond. 

Cooling of the sample is achieved through an He exchange gas between the sample and the cold plate of the cryostat maintained at a typical temperature of 3-3.5 K, which also cools down the superconducting magnet. The top plate of the positioner stack holding the sample contains a resistive heater and a calibrated temperature sensor (Lakeshore Cernox CX-1050-CU-HT-1.4L). In the experiments presented here, the sensor temperature was typically $T_{\rm sensor}=4.0-4.5$~K in the absence of any deliberate heating, rising to $T_{\rm sensor}=4.5-5.0$~K during NV imaging with 50 mW of laser power. For temperature-dependent measurements, the resistive heater was driven by a temperature controller (Lakeshore 335), see further details in section \ref{sec:temp}.    

\section{NV measurements} \label{sec:NVmeas}

The magnetic measurements reported in the main text involved performing optically-detected magnetic resonance (ODMR) spectroscopy of the NV layer. We used a pulsed ODMR sequence with a $10~\mu$s laser pulse and a 100-150 ns microwave $\pi$ pulse, repeated 3000 times for each microwave frequency to match the 30 ms exposure time of the camera. The spectra were normalised by taking a frame without microwave after each frame with microwave. A single frequency sweep typically takes of the order of a second. The sweep is repeated thousands of times to improve the signal to noise ratio, corresponding to a total acquisition time per magnetic image from tens of minutes to several hours.

All the NV measurements shown in the main text are taken in a small bias magnetic field $B_{\rm NV}^{\rm bias}=5$~mT aligned with the [111] direction of the diamond crystal, which forms a $54.7^\circ$ angle with the $z$ axis. The $z$ axis is defined as the normal to the diamond surface and also corresponds to the $c$ axis of the VI$_3$ flakes. This bias magnetic field is large enough to clearly separate the aligned NV centres from the other three orientation families in the ODMR spectrum, but is small enough to have a negligible effect on the magnetization of the VI$_3$ flakes given the much larger coercive field and anisotropy field~\cite{Yan2019}. The magnetic images therefore show the remanent magnetization of the sample.    

\begin{figure*}[htb!]
	\begin{center}
		\includegraphics[width=0.9\textwidth]{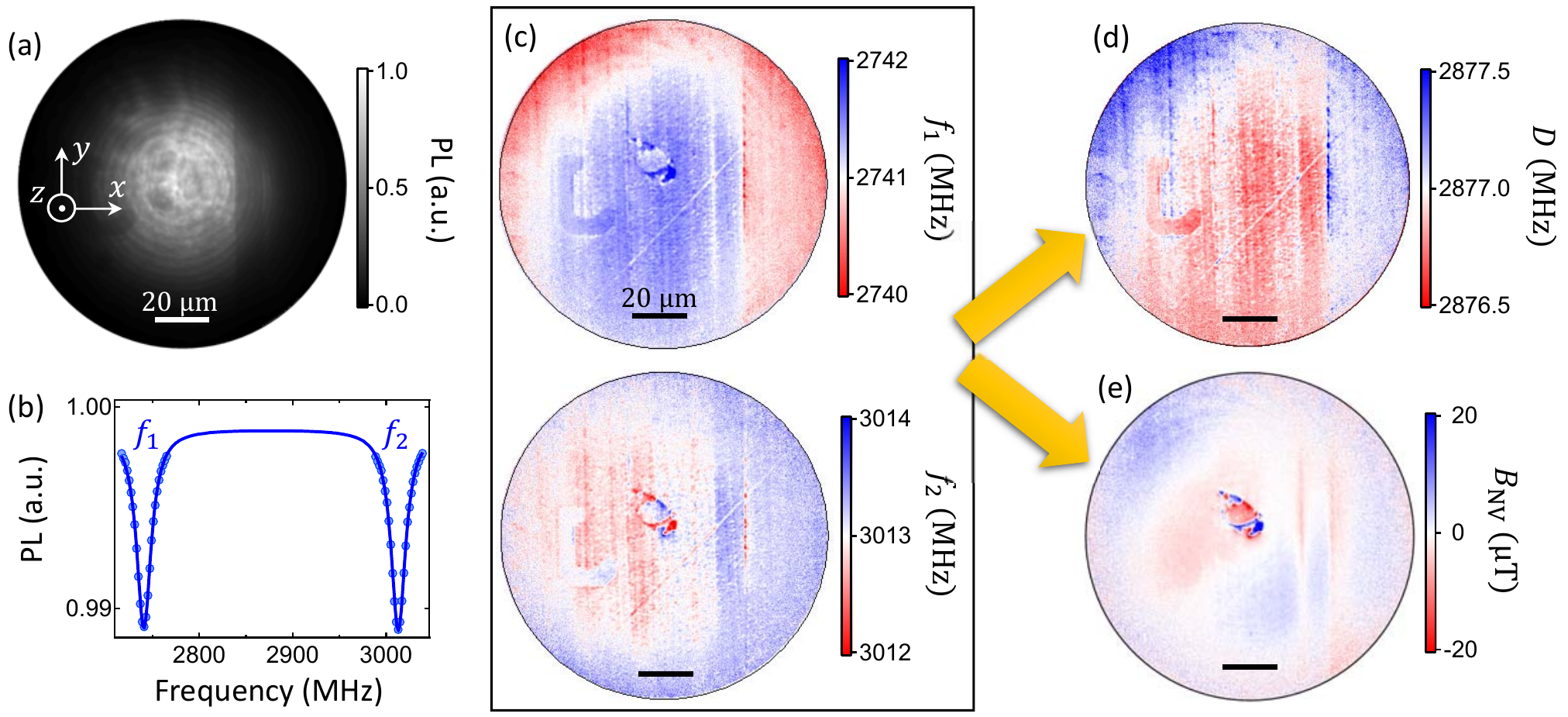}
		\caption{\textbf{Widefield magnetic imaging with NV centres.} (a) PL image of the NV layer under sample \#2. (b) Example ODMR spectrum from a given pixel. The solid line is a fit with two Lorentzian peaks at free frequencies $f_1$ and $f_2$. (c) Maps of the frequencies $f_1$ and $f_2$. (d) Map of the zero-field splitting parameter $D=(f_1+f_2)/2$ calculated from (c). (e) Map of the magnetic field projection $B_{\rm NV}=(f_2-f_1)/2\gamma_{\rm NV}$ calculated from (c). A constant offset of 4.852 mT was subtracted in the plotted map.
		}
		\label{FigSI_ODMR}
	\end{center}
\end{figure*}

Example ODMR data from sample \#2 are shown in Fig.~\ref{FigSI_ODMR}. The PL image in Fig.~\ref{FigSI_ODMR}a reveals the laser spot size which has a $\approx50~\mu$m waist. This provides sufficient illumination across the $125~\mu$m field of view of the camera. An example ODMR spectrum from a given pixel from this image is shown in Fig.~\ref{FigSI_ODMR}b. The two dips visible in the spectrum correspond to the $|0\rangle\rightarrow|-1\rangle$ and $|0\rangle\rightarrow|+1\rangle$ electron spin resonances of the NV centres aligned with the magnetic field, with transition frequencies $f_1$ and $f_2$, respectively. Fitting each spectrum with a sum of two Lorentzian functions with free frequencies, amplitudes and widths give the two frequency maps shown in Fig.~\ref{FigSI_ODMR}c. These frequencies depend on the magnetic field according to $f_1=D-\gamma_{\rm NV}B_{\rm NV}$ and $f_2=D+\gamma_{\rm NV}B_{\rm NV}$, where $D$ is the zero-field splitting and $\gamma_{\rm NV}=28.035(3)$~GHz/T is the NV gyromagnetic ratio~\cite{Doherty2013,Rondin2014}. The parameter $D$ depends on the local strain and is found to vary by 100's of kHz across the field of view due to polishing damage as well as strain induced by the Al grid, as highlighted in Fig.~\ref{FigSI_ODMR}d which plots the sum of the two measured frequencies, $D=(f_1+f_2)/2$. This strain has an effect of comparable magnitude to the magnetic field from the VI$_3$ flakes, however it is efficiently normalised out by taking the difference of the measured frequencies, $B_{\rm NV}=(f_2-f_1)/2\gamma_{\rm NV}$. 

The resulting magnetic field map is shown in Fig.~\ref{FigSI_ODMR}e, where the mean value has been subtracted to isolate the contribution of the VI$_3$ flakes. In this example the mean value was 4.852 mT. The difference with the applied bias field of $B_{\rm NV}^{\rm bias}=5.000$~mT is attributed to residual magnetic fields emanating from the superconducting coils. Aside from the magnetic signal from the VI$_3$ flakes (near the centre of the image), the $B_{\rm NV}$ map shown in Fig.~\ref{FigSI_ODMR}e exhibits long-range background variations of the order of $10~\mu$T. These variations indicate small biases in the fit frequency positions across the field of view, presumably related to variations in laser intensity and/or microwave field amplitude affecting the NV spin contrast during the microwave frequency sweep. These long-range variations give rise to low frequency components in the reconstructed magnetization maps, which are removed through a background subtraction algorithm (see section \ref{sec:MagRecon}).

\section{Magnetization reconstruction} \label{sec:MagRecon}

The magnetic field is measured along a known direction defined by the unit vector ${\bf u}_{\rm NV}=(1,1,1)/\sqrt{3}$, using the $xyz$ reference frame defined in Fig.~\ref{FigSI_ODMR}a. That is, one measures the projection $B_{\rm NV}={\bf B}\cdot{\bf u}_{\rm NV}$ where ${\bf B}$ is the magnetic field at the NV layer, which is the sum of the bias field ${\bf B}_{\rm NV}^{\rm bias}=B_{\rm NV}^{\rm bias}{\bf u}_{\rm NV}$ and the stray field emanating from the magnetic VI$_3$ flakes, ${\bf B}_{\rm stray}$. In the case of an ultrathin magnetic sample of magnetization ${\bf M}=M(x,y){\bf u}_M$ where the direction ${\bf u}_M$ is constant and known, it is possible to uniquely reconstruct (up to some offsets) the amplitude $M(x,y)$ from the magnetic field map $B_{\rm NV}(x,y)$~\cite{Casola2018,Thiel2019}. Here, the VI$_3$ flakes are expected to be magnetized perpendicular to their plane, i.e. along the $c$ axis~\cite{Tian2019}, and so we take ${\bf u}_M=(0,0,1)$ and solve for the amplitude $M_z(x,y)$. In the Fourier plane $(k_x,k_y)$, the relationship between the field $b_{\rm NV}(k_x,k_y,z)$ at a standoff distance $z$ above the flakes and the magnetization $m_z(k_x,k_y)$ reads
\begin{eqnarray} \label{eq:m_to_b_singlez}
b_{\rm NV}(k_x,k_y,z)={\bf u}_{\rm NV}\cdot(-ik_x,-ik_y,k)\frac{\mu_0}{2}e^{-kz}m_z(k_x,k_y)t
\end{eqnarray}
where $k=\sqrt{k_x^2+k_y^2}$, $\mu_0$ is the vacuum permeability and $t$ is the flake thickness. To take into account the finite thickness of the NV layer (see Fig.~\ref{FigSI_diamonds}), we integrate Eq.~\ref{eq:m_to_b_singlez} over the range $z\in[z_{\rm NV}-t_{\rm NV}/2,z_{\rm NV}+t_{\rm NV}/2]$ where $z_{\rm NV}$ is the mean standoff distance and $t_{\rm NV}$ is the NV layer thickness, which gives
\begin{eqnarray} \label{eq:m_to_b}
\bar{b}_{\rm NV}(k_x,k_y)={\bf u}_{\rm NV}\cdot(-ik_x,-ik_y,k)\frac{\mu_0}{2}e^{-kz_{\rm NV}}\frac{\sinh(kt_{\rm NV}/2)}{kt_{\rm NV}/2}m_z(k_x,k_y)t
\end{eqnarray}
where $\bar{b}_{\rm NV}$ denotes the average field over the NV layer, assuming a uniform distribution of NVs within the interval define above.  

For $(k_x,k_y)=(0,0)$, the right-hand-side term in Eq.~\ref{eq:m_to_b} vanishes and so $m_z(0,0)$ is not defined. In real space this corresponds to an undetermined global offset, which we will set by requiring the magnetization to be null far from the flakes. For any non-zero $k$-vector, Eq.~\ref{eq:m_to_b} can be inverted to give 
\begin{eqnarray} \label{eq:b_to_m}
m_z(k_x,k_y)t={\cal T}(k_x,k_y)\bar{b}_{\rm NV}(k_x,k_y)
\end{eqnarray}
where 
\begin{eqnarray} \label{eq:T}
{\cal T}(k_x,k_y)=\frac{kt_{\rm NV} e^{kz_{\rm NV}}}{\mu_0\sinh(kt_{\rm NV}/2){\bf u}_{\rm NV}\cdot(-ik_x,-ik_y,k)}
\end{eqnarray}
The Fourier transform of $m_z(k_x,k_y)t$ corresponds to an areal magnetization (in units of A or $\mu_B/{\rm nm}^2$) and is simply denoted as $M_z$ in the main text. The $M_z$ maps shown in the main text were produced using $z_{\rm NV}=290$~nm and $t_{\rm NV}=100$~nm, based on Fig.~\ref{FigSI_diamonds}a,b. We note that the pixel size in our images (250 nm) is close to $z_{\rm NV}$, and so the $e^{kz_{\rm NV}}$ factor in Eq.~\ref{eq:T} remains close to one even at the highest spatial frequencies. Nevertheless, to prevent any noise amplification, the ${\cal T}(k_x,k_y)$ matrix was passed through a Hanning low-pass filter with cut-off frequency $k_c=2/z_{\rm NV}$, as done in Ref.~\cite{Thiel2019}.  

\begin{figure*}[htb!]
	\begin{center}
		\includegraphics[width=0.6\textwidth]{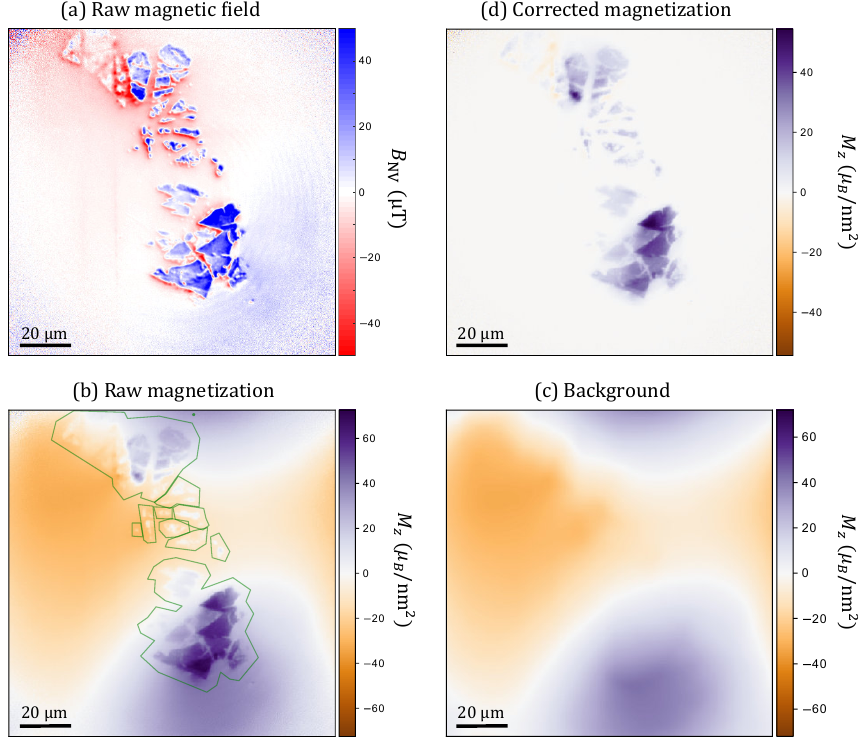}
		\caption{\textbf{Magnetization reconstruction and background subtraction.} (a) Raw $B_{\rm NV}$ map of sample \#1 after a simple offset subtraction. Fig. 1c of the main text corresponds to the same map after a plane subtraction. (b) Raw $M_z$ reconstructed from (a) using Eq. \ref{eq:b_to_m}. The $k=0$ component was set to zero. (c) Background obtained from (b) by removing the regions corresponding to physical magnetic features displayed as green polygons in (b), see details in the text. (d) Background-corrected $M_z$ map obtained by subtracting (c) to (b).
		}
		\label{FigSI_MzCorrection}
	\end{center}
\end{figure*}

In the raw $M_z$ maps, low-frequency variations are typically observed in the background (see Fig.~\ref{FigSI_MzCorrection}b for sample \#1), which are due to low-frequency components in the raw $B_{\rm NV}$ maps (Fig.~\ref{FigSI_MzCorrection}a) caused by small systematic ODMR fitting errors varying across the field of view (see section \ref{sec:NVmeas}). These variations are not physical since the magnetization should vanish outside the flakes. To allow a more accurate reading of the magnetization of the flakes, we applied a background subtraction procedure as follows. First, the regions where physical flakes are present in the field of view are defined based on the optical images and the presence of sharp magnetic features in the $B_{\rm NV}$ map, indicated by green polygons in Fig.~\ref{FigSI_MzCorrection}b. The background is then estimated by removing these regions from the raw $M_z$ map, replacing them by a linear interpolation, and applying a Gaussian convolution to produce a smooth background with only low-frequency components (Fig.~\ref{FigSI_MzCorrection}c). Subtracting this background from the raw $M_z$ map gives the corrected $M_z$ map (Fig.~\ref{FigSI_MzCorrection}d). All the $M_z$ maps shown in the main text were obtained via this procedure.

\section{Direct comparison with theoretical stray field} \label{sec:Comparison}

\begin{figure*}[b!]
	\begin{center}
		\includegraphics[width=0.8\textwidth]{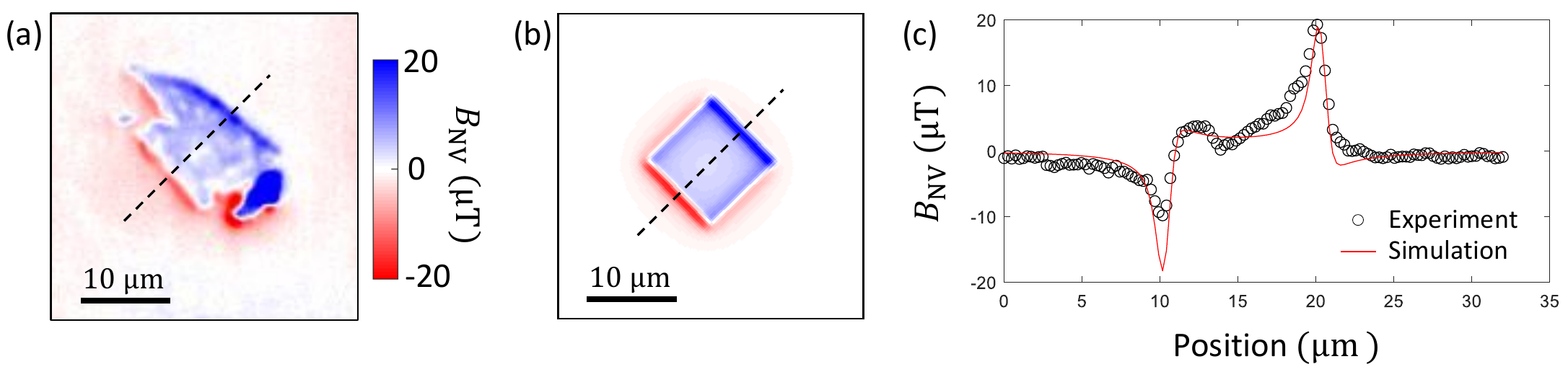}
		\caption{\textbf{Comparison of measured stray field with simple model.} (a) $B_{\rm NV}$ map of sample \#2. (b) Calculated $B_{\rm NV}$ map for a $10~\mu$m~$\times~10~\mu$m flake with a uniform magnetization $M_z=5~\mu_B/{\rm nm}^2$. The standoff is taken to be $z_{\rm NV}=290$~nm, and we convolved the result with a 2D Gaussian function with a full-width at half maximum of 700 nm to model the finite spatial resolution of the measurement. (c) Line cuts taken along the black dashed lines shown in (a,b).
		}
		\label{FigSI_Comparison}
	\end{center}
\end{figure*}

To verify that the $M_z$ values obtained via the reconstruction method are sound, we directly compared the magnetic field data with the expected field for a simplified geometry. We consider the data for sample \#2 (Fig.~\ref{FigSI_Comparison}a) and model the flake as a square-shaped 2D magnetic element $10~\mu$m~$\times~10~\mu$m in size, with a uniform magnetization $M_z$. The stray field outside such a structure can be readily computed, for instance using the analytical solutions from Ref.~\cite{Rondin2012}. The calculated stray field ${\bf B}$, at a given standoff distance $z_{\rm NV}$, is then projected along the NV axis ${\bf u}_{\rm NV}$ to give $B_{\rm NV}$. To model the finite spatial resolution of the measurement, we convolved the raw $B_{\rm NV}$ map with a 2D Gaussian function with a full-width at half maximum of 700 nm. The result is shown in Fig.~\ref{FigSI_Comparison}b for a magnetization $M_z=5~\mu_B/{\rm nm}^2$ and a standoff $z_{\rm NV}=290$~nm, reproducing the overall pattern seen in the experiment (Fig.~\ref{FigSI_Comparison}a). Line cuts through the flake (Fig.~\ref{FigSI_Comparison}c) confirm the broad agreement between experiment (data points) and simulation (solid line), indicating that indeed the magnetization of this 3-layer VI$_3$ flake is of the order of $5~\mu_B/{\rm nm}^2$, much smaller than expected from measurements of bulk VI$_3$ ($\approx15~\mu_B/{\rm nm}^2$ for 3 layers). 

We note that the non-trivial shape and non-uniform magnetization of the flakes prevent an accurate fitting of the $B_{\rm NV}$ data to directly extract $M_z$ as was done in Ref.~\cite{Thiel2019}, as highlighted by the discrepancies between experiment and simulation in Fig.~\ref{FigSI_Comparison}c. In this case, it is thus preferable to use the reconstructed $M_z$ map which automatically accounts for arbitrary shapes and magnetization patterns. The only downside of the $M_z$ reconstruction method is that it may introduce truncation artefacts when computing the Fourier transform of $B_{\rm NV}$, however this was found to be negligible in most cases given that our field of view is much larger than the size of the flakes studied.

\section{Temperature dependence of the magnetization} \label{sec:temp}
       
\begin{figure*}[b!]
	\begin{center}
		\includegraphics[width=1\textwidth]{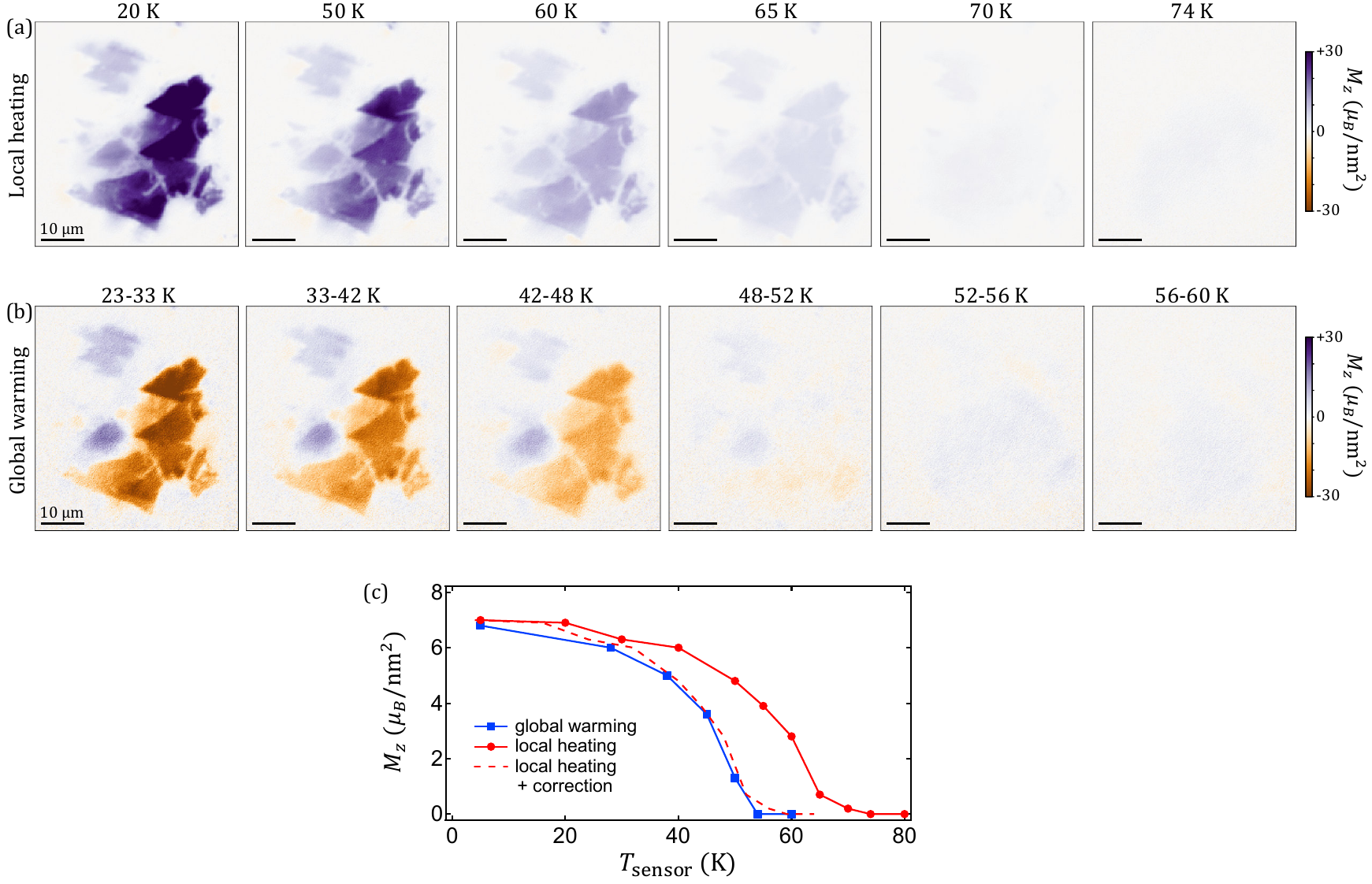}
		\caption{\textbf{Magnetization maps vs temperature.} (a) $M_z$ maps of sample \#1 at various temperatures obtained by local heating with a resistive heater. The temperature measured by a nearby sensor, $T_{\rm sensor}$, is indicated above each image. (b) $M_z$ maps recorded during a slow global warming of the system (see text), where the sample temperature ($T_{\rm sample}$) should coincide with the sensor temperature ($T_{\rm sensor}$). The temperature range during the measurement (start-end temperature as measured by the sensor) is indicated above each image. (c) Magnetization of an 8-layer flake as a function of $T_{\rm sensor}$ with local heating and global warming. Multiplying the $T_{\rm sensor}$ scale by a factor 0.8 brings the local heating curve (dashed line) close to the global warming curve, providing a way to estimate $T_{\rm sample}$ when using the local heating method.         
		}
		\label{FigSI_temperature}
	\end{center}
\end{figure*}

To vary the sample temperature, $T_{\rm sample}$, two strategies can be employed. The most natural one is to use a resistive heater to regulate the temperature recorded by the nearby sensor, $T_{\rm sensor}$. However, we found that this local heating generally produces significant temperature gradients across the system due to poor thermal coupling between the sample (i.e. the VI$_3$ flakes) and the heater/sensor. This is evidenced by the fact that the onset of magnetism occurs at a sensor temperature $T_{\rm sensor}\approx70$~K (Fig.~\ref{FigSI_temperature}a), which is larger than the Curie temperature of bulk VI$_3$ ($T_c\approx50$~K)~\cite{Tian2019}. This means that the VI$_3$ flakes, which are exposed to the cold He exchange gas, are colder (by about 20 K) than the heater/sensor plate. The poor thermal coupling between the VI$_3$ flakes and the heater/sensor plate is likely a result of the low thermal conductivity of some of the elements (e.g. the glass coverslip supporting the diamond) and poor thermal contacts within the PCB/coverslip/diamond assembly (see section~\ref{sec:setup}).   

An alternative strategy to vary $T_{\rm sample}$ is to reduce the overall cooling power of the cryostat. This is achieved by reducing the pressure of the He exchange gas thermally coupling the cold plate with the vacuum tube housing the sample, which is itself filled with He exchange gas (room temperature pressure of 10-100 mbar). By reducing the pressure in the outer space while keeping the inner pressure high, we can thus increase the sample temperature while minimising temperature gradients inside the tube since there are no local sources of heating or cooling. In that case, the sensor temperature is a good estimate of the sample temperature, $T_{\rm sensor}\approx T_{\rm sample}$, as confirmed by the fact that the onset of magnetism is seen at $T_{\rm sensor}\approx T_c$ (Fig.~\ref{FigSI_temperature}b). The data plotted in Fig.~2d of the main text were obtained using this method. Namely, starting from high cooling power where $T_{\rm sensor}<5$~K, we decreased the He pressure in the outer space causing $T_{\rm sensor}$ to slowly increase up to 100 K over the course of a few hours. NV measurements were taken throughout this slow increase and the temperature monitored. The $M_z$ maps thus obtained are shown in Fig.~\ref{FigSI_temperature}b, also displaying the temperature range during each measurement, that is, the starting and final temperatures. In Fig.~2d of the main text, the temperature was taken as the mean value of this temperature range. 

The $M_z$ vs $T_{\rm sensor}$ curves obtained via these two methods are shown in Fig.~\ref{FigSI_temperature}c for an 8-layer flake. We notice that the local heating curve seems to be roughly a stretched version of the global warming curve. That is, multiplying the $T_{\rm sensor}$ scale by a factor 0.8 brings the local heating curve close to the global warming curve. In Fig. 3c of the main text, we used this correction to estimate the sample temperature, where $T_{\rm sensor}$ was regulated by local heating to allow longer measurements. 

\section{Domain reversal}

In Fig. 3 of the main text, we imaged the magnetic switching of several samples. Here we give further details on the experiments and analysis. The samples were initially magnetized by applying a magnetic field of 1~T in the $+z$ direction for a few minutes, at the base temperature ($T_{\rm sensor}\approx5$~K). Magnetic field pulses of increasing amplitude $B_p$ were then applied in the $-z$ direction to switch the sign of the magnetization, i.e. $B_p=-0.1$~T, $-0.2$~T, $-0.3$~T etc. The pulses were applied by ramping up the field at a rate of 50~mT/s until $B_p$ is reached, then maintaining $B_p$ for about 10 seconds, and ramping down to zero at a rate of $-50$~mT/s. After each step, $B_{\rm NV}$ maps were recorded under a small bias magnetic field ($B_{\rm NV}^{\rm bias}=5$~mT) and converted to $M_z$ maps. 

When a flake has not switched ($M_z>0$) after a pulse of $B_p=-0.2$~T but has ($M_z<0$) after a pulse of $B_p=-0.3$~T, for instance, we assign to this flake a switching field (i.e the coercive field in this case) of $H_c=-0.25\pm0.05$~T, which is the half-way point. Because magnetization switching is a random process, a more accurate estimate could in principle be obtained by repeating the process many times and building a histogram of the switching field. Here, given the long acquisition times involved, we simply verified that repeating the process once (re-magnetizing the flakes in the $+z$ direction and applying pulses in the $-z$ direction) gave a similar switching field, typically within 0.1~T. We also verified that the switching field is symmetric, that is, switching the flake back in the $+z$ direction requires the same field amplitude (within 0.1 T) just of opposite sign.     
  
\begin{figure*}[htb!]
	\begin{center}
		\includegraphics[width=1\textwidth]{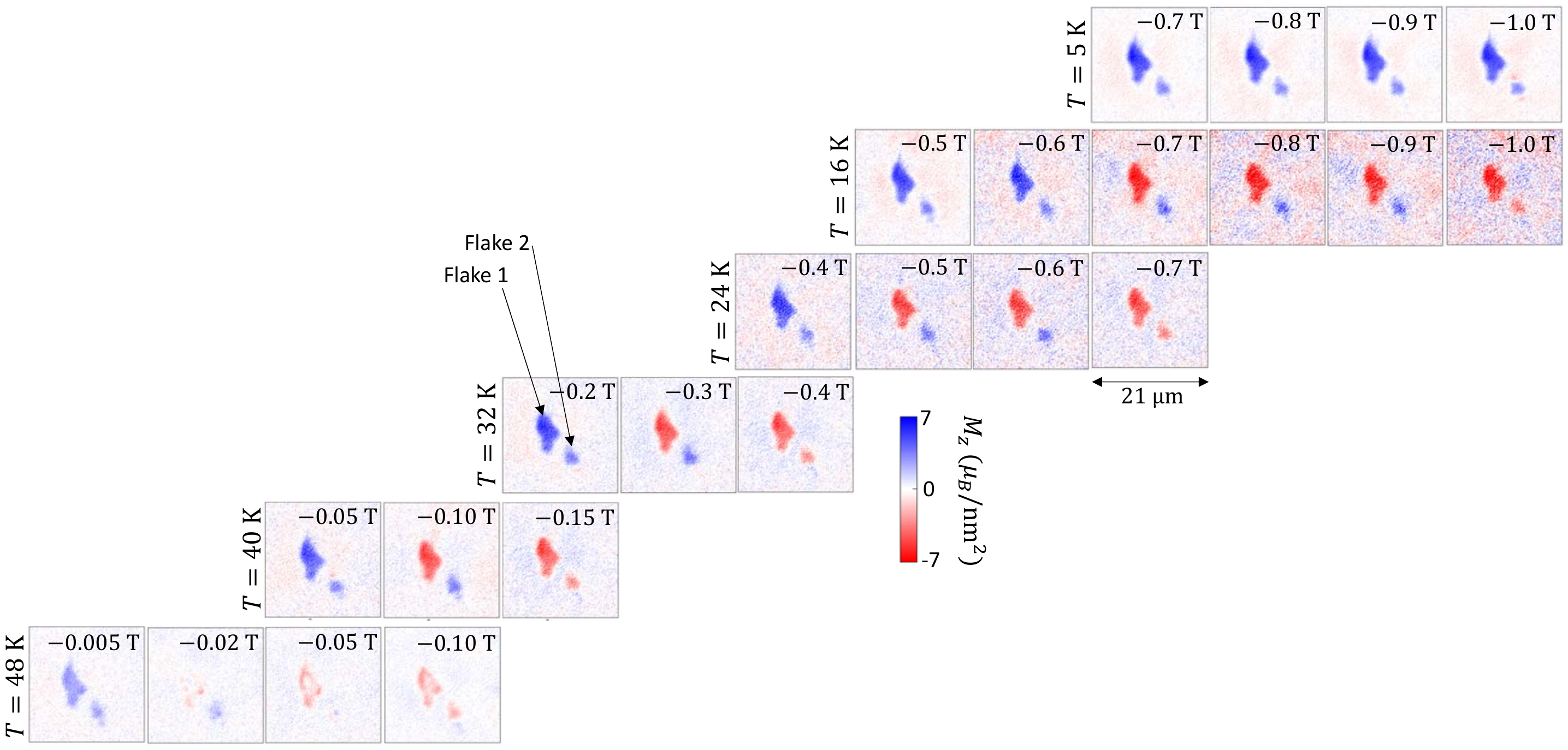}
		\caption{\textbf{Magnetization reversal vs temperature.} $M_z$ maps of sample \#3 after applying an increasingly large magnetic field pulse in the $-z$ direction, starting from the flakes magnetized in the $+z$ direction. Each row corresponds to a different sample temperature. These maps were used to estimate the coercive field $H_c$ in Fig. 3c of the main text. 
		}
		\label{FigSI_Hc_vs_T}
	\end{center}
\end{figure*}

For sample \#3, the process was repeated at different temperatures to construct the graph in Fig. 3c of the main text. For these measurements, the temperature was set using local heating and the measured sensor temperature was corrected to obtain an estimate ($\pm5$~K) of the sample temperature as explained in section~\ref{sec:temp}. For each temperature, we re-initialized the magnetic state by applying 1~T in the $+z$ direction before applying the pulses in the $-z$ direction. The full series of $M_z$ maps versus temperature and pulse amplitude are shown in Fig.~\ref{FigSI_Hc_vs_T}. From these maps, for each flake (labelled 1 and 2) and each temperature, the switching field $H_c$ was determined as described above.

In the Stoner-Wohlfarth model~\cite{Coey2010}, the coercive field of a thin magnetic element with perpendicular anisotropy is given by the anisotropy field, $H_c=\frac{2K_{\rm eff}}{M_s}$. Here $K_{\rm eff}=K_1-\mu_0M_s^2/2$ is the effective anisotropy constant, where $K_1$ is the magneto-crystalline anisotropy constant and the second term is the demagnetizing contribution which reduces the effective anisotropy. This coercive field corresponds to the nucleation field of the coherent reversal mode~\cite{Coey2010}. In Ref.~\cite{Yan2019}, the anisotropy constant of bulk VI$_3$ was determined to be $K_1=37$~MJ/m$^3$ at 5 K, with an approximately linear dependence with temperature where the anisotropy is reduced to $K_1=5$~MJ/m$^3$ at 48 K. Using these values together with a spontaneous magnetization of $M_s=70$~kA/m (corresponding to 1 $\mu_B$/f.u.), we obtain a coercive field of $H_c\approx1.0$~T at 5 K. Using the temperature dependence of $M_s$ from Ref.~\cite{Yan2019}, we get $H_c\approx0.2$~T at 48 K, with an approximately linear decrease between these temperatures. This is the origin of the dashed line plotted in Fig. 3c of the main text. 

In the case of sample \#2, we observed a slightly different behaviour compared to the other samples. By initializing the sample as before, i.e. by applying 1~T at 5 K to magnetize the flakes in the $+z$ direction, reversed domains were formed in the 3-layer flake at a field as low as $-0.3$~T, as shown in Fig.~\ref{FigSI_reversal}a. Upon increasing the field amplitude, the domains grew in size but full reversal of the flake was achieved only at $-0.8$~T, with clear indication of domain wall pinning in between. However, a second initialization, this time by applying 1~T while cooling down from above $T_c$, gave a different result, shown in Fig.~\ref{FigSI_reversal}b. Here, there is no switching of the majority of the flake even at $-1$~T, implying that the switching field is much larger than the domain wall depinning fields, similar to the other samples. One possible explanation is that in the first case, the flake was not properly initialized, leaving pockets of reversed magnetization that acted as nucleation points. We also notice that the magnetization is slightly larger in the second case compared to the first one, $\approx8~\mu_B/$nm$^2$ instead of $\approx5~\mu_B/$nm$^2$ for the 3-layer flake. This may be further evidence of incomplete initialization in the first case, although we stress that the second measurement was performed several weeks after the first one and so other factors may have contributed to the discrepancy (e.g., a slightly different base temperature).

\begin{figure*}[htb!]
	\begin{center}
		\includegraphics[width=1\textwidth]{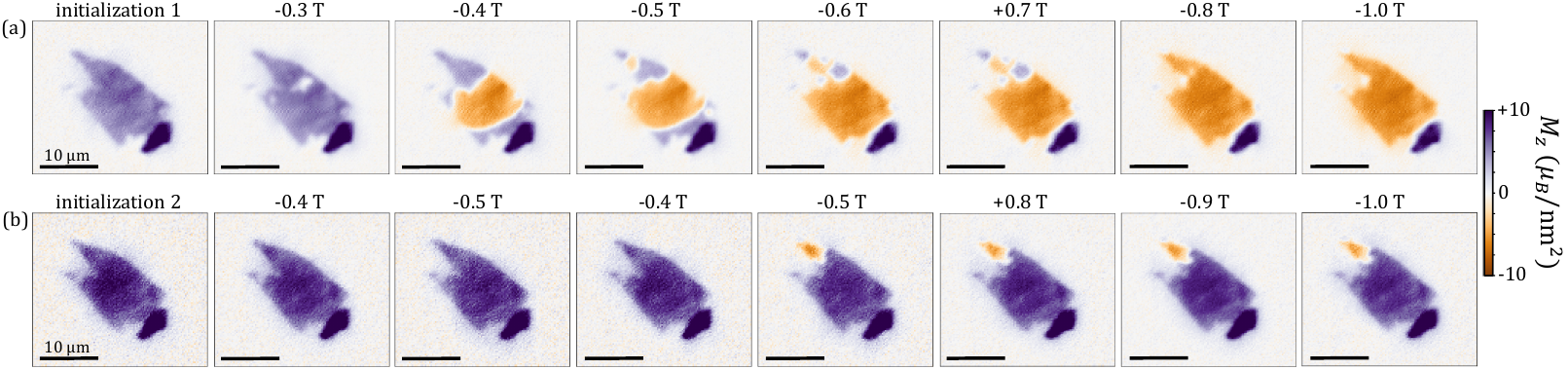}
		\caption{\textbf{Additional data of magnetization reversal.} $M_z$ maps of sample \#2 as a function of the magnetic field pulse amplitude, $B_p$, starting from two different initial states. In the first case (a), the sample was initialized by applying 1~T at 5 K following a zero field cooling, while in the second case (b) the field was applied during cooling to ensure complete magnetization.
		}
		\label{FigSI_reversal}
	\end{center}
\end{figure*}

\section{Initial magnetization and depinning field}

In Fig. 4 of the main text, we imaged the evolution of sample \#1 during the initial magnetization. Here we give further details on the experiments and analysis. The sample was initially prepared in a virgin state by heating it above $T_c$ using local heating, demagnetizing the superconducting magnet (by alternating positive and negative sign while decreasing the amplitude) to remove any residual magnetic field, and cooling the sample down to the base temperature ($T_{\rm sensor}\approx5$~K) under no applied magnetic field. Following this zero field cooling (ZFC), magnetic field pulses ($\sim10$~second long) of increasing amplitude $B_p$ were applied in the $+z$ direction to progressively magnetize the sample. After each step, $B_{\rm NV}$ maps were recorded under a small bias magnetic field ($B_{\rm NV}^{\rm bias}=5$~mT) and converted to $M_z$ maps. 

To construct the depinning field histograms in Fig. 4c of the main text, the number of changes induced by each magnetic field pulse were counted. Here a change was defined by either the jump of a domain wall segment or, where domain walls are not clearly resolved because the domain size is close to the spatial resolution, by the change of size or intensity of the domain. The histograms for the 4L, 5L and 8L flakes were obtained from the image series in Fig. 4a of the main text, whereas the histogram for the `thick' flakes was obtained by analysing the region indicated by the dashed box in Fig. 4b of the main text, which corresponds to flakes 10-20 nm in thickness.

Additional initial magnetization data are shown in Figs.~\ref{FigSI_initial_mag}a and \ref{FigSI_initial_mag}b for samples \#2 and \#3, showing similar depinning fields to sample \#1 in the range 0.2-0.3~T. 

\begin{figure*}[htb!]
	\begin{center}
		\includegraphics[width=1\textwidth]{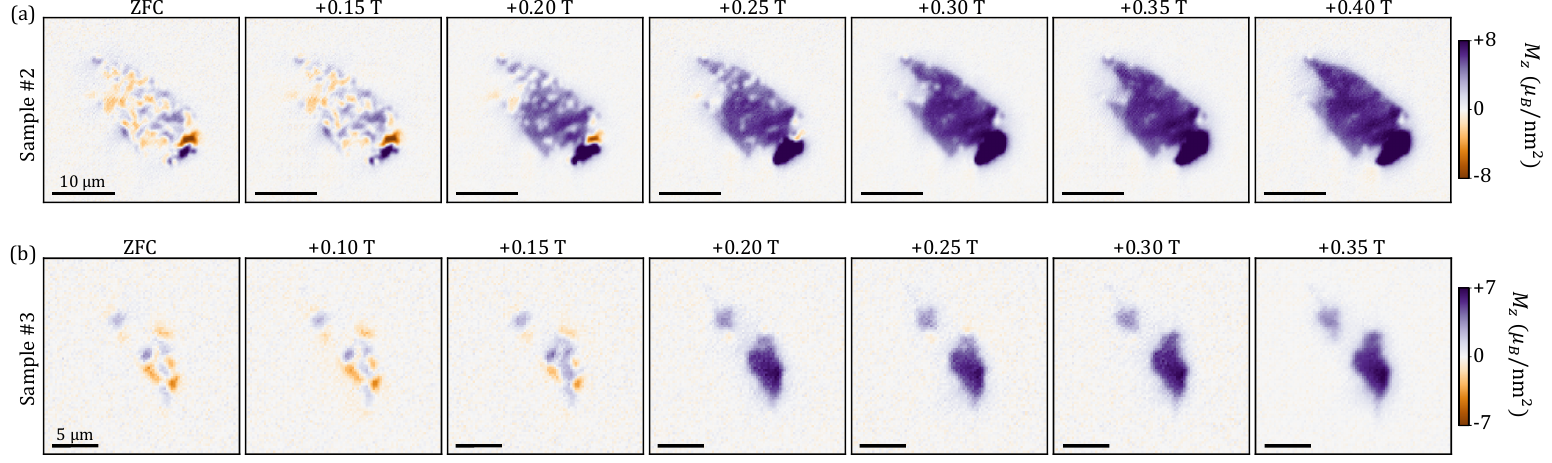}
		\caption{\textbf{Additional data of initial magnetization.} $M_z$ maps of (a) sample \#2 and (b) sample \#3 as a function of the magnetic field pulse amplitude, $B_p$, starting from the virgin state (zero-field cooling, ZFC). 
		}
		\label{FigSI_initial_mag}
	\end{center}
\end{figure*}

In magnetic films with perpendicular anisotropy, the energy per unit domain wall area is $\sigma_0=4\sqrt{AK_{\rm eff}}$ where $A$ is the exchange interaction and $K_{\rm eff}$ is the effective anisotropy constant. In a 1D model, the domain wall energy at a pinning site can be expressed as $\sigma_{\rm pin}=(1-\alpha)\sigma_0$ where $\alpha$ is a factor ranging between 0 (no pinning) and 1 (corresponding to a void where $A=K_{\rm eff}=0$). The local minimum in the energy versus position has a minimum width given by the domain wall width, $\Delta=\sqrt{A/K_{\rm eff}}$. In a perpendicular magnetic field of amplitude $H$, the energy versus position profile is tilted by an amount $2M_sH$ per unit of displacement. The depinning field $H_d$ is approximately given by the condition $\alpha\sigma_0=2M_sH_d\Delta$, which gives $H_d\sim\alpha\frac{2K_{\rm eff}}{M_s}$. The maximum depinning field, corresponding to a void, is therefore $H_d^{\rm max}\sim\frac{2K_{\rm eff}}{M_s}$, which also corresponds to the coercive field $H_c$ in the Stoner-Wohlfarth model~\cite{Coey2010}. In a flake comprising $N$ layers, a void in only one of the layers will give a pinning factor $\alpha=1/N$ hence $H_d\sim\frac{2K_{\rm eff}}{NM_s}$. The pinning fields measured in the thin flakes of sample \#1 are in broad agreement with this expression, which predicts $H_d\sim0.25$~T for a 4-layer flake, for instance. This suggests that the strong pinning sites observed in the images may be due to defects such as cracks extending through a single atomic layer. The fact that the depinning field decreases with increasing thickness (increasing $N$) as observed in Fig.~4c of the main text supports this picture, although the nonlinear scaling indicates that for thicker flakes the defects may extend through several layers.

\section{Effect of laser heating}

Even though the VI$_3$ flakes are not directly illuminated by the laser and are somewhat thermally decoupled from the laser-illuminated diamond (see section \ref{sec:diamonds}), some amount of heating may still reach the VI$_3$ flakes. To quantify this effect, we recorded images at varying laser powers, $P_{\rm Laser}$. The results are presented in Fig.~\ref{FigSI_laser} for sample \#3 with laser powers (entering the cryostat) ranging from 1 mW to 35 mW. The $B_{\rm NV}$ and $M_z$ maps (Fig.~\ref{FigSI_laser}a,b) are qualitatively similar across the different laser powers, and line cuts reveal only small variations (Fig.~\ref{FigSI_laser}c). To quantify these variations, we integrate the $M_z$ maps over the two flakes (dotted box in Fig.~\ref{FigSI_laser}b) to obtain the total magnetic moment. This is plotted as a function of $P_{\rm Laser}$ in Fig.~\ref{FigSI_laser}d, revealing a small decrease with increasing laser power, where the moment at 35 mW (10 mW) is about 10\% (8\%) smaller than at 1 mW. Thus, we conclude that while the laser contributes to slightly reduce the measured magnetization (either due to local heating or to laser-dependent biases in the ODMR fitting), it cannot be solely responsible for the discrepancy between the magnetization of our few-layer flakes and that of bulk VI$_3$ observed in Fig.~2c of the main text.  

In fact, we postulate that laser-induced heating is mostly negligible as far as the magnetic properties of VI$_3$ are concerned, based on the following observations. First, the images in Fig.~\ref{FigSI_temperature} (recorded with $P_{\rm Laser}=50$~mW) show that the onset of magnetism occurs when the sensor temperature is $T_{\rm sensor}\approx50$~K consistent with the Curie temperature of bulk VI$_3$~\cite{Tian2019}. If there was a significant laser heating at the sample location compared to the sensor location, then the apparent onset would occur at a lower $T_{\rm sensor}$. Second, the coercive field $H_c$ is expected to decrease when the temperature is increased~\cite{Yan2019}. In Fig. 3 of the main text, we found that $H_c$ is of the order of 1 T for most flakes without active heating (apart from the laser), consistent with bulk VI$_3$ at 2 K~\cite{Yan2019}. Thus, the measured $H_c$ is consistent with the sample being at a temperature close to the base temperature of $T_{\rm sensor}\approx5$~K. Therefore, given the lack of evidence of any major effect of the laser on the magnetic properties of VI$_3$, all the data reported in the paper were acquired with a relatively high laser power, between 10 mW and 50 mW, unless specified otherwise.

\begin{figure*}[htb!]
	\begin{center}
		\includegraphics[width=0.8\textwidth]{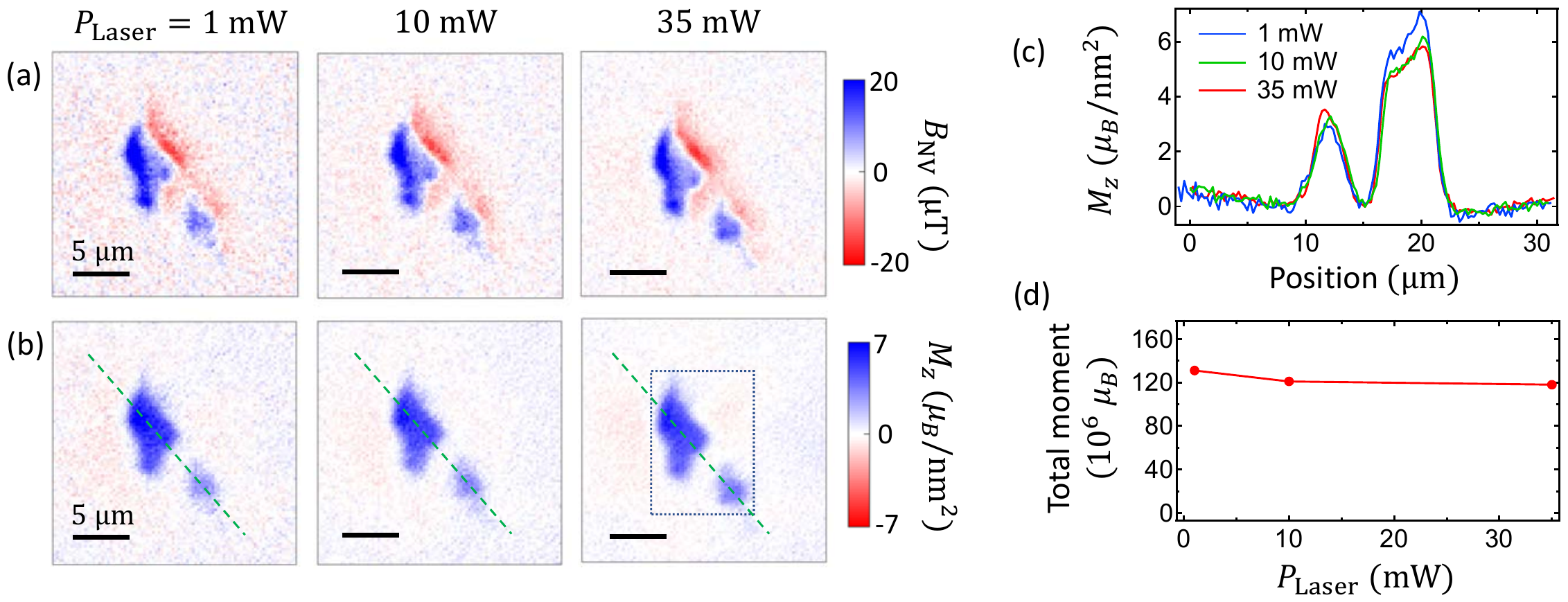}
		\caption{\textbf{Effect of laser heating.} (a) $B_{\rm NV}$ maps of sample \#3 obtained with different laser powers $P_{\rm Laser}$. (b) Corresponding $M_z$ maps. (c) Line cuts of $M_z$ taken along the green dashed line shown in (b). (d) Total magnetic moment obtained by integrating the dotted box shown in (b), as a function of $P_{\rm Laser}$. 
		}
		\label{FigSI_laser}
	\end{center}
\end{figure*}

\section{Effect of the bias magnetic field}

All NV measurements were performed in a small bias magnetic field of $B_{\rm NV}^{\rm bias}=5$~mT, revealing the remanent magnetic state of the samples. In the absence of domains, this remanent magnetization corresponds to the spontaneous magnetization $M_s$. However, because the local magnetization in our experiments was found to be somewhat lower than the bulk $M_s$, a legitimate question is whether small domains (below our spatial resolution) may spontaneously form at low field, which would decrease the apparent (averaged) magnetization. To address this question, we measured sample \#1 again but in a larger bias magnetic field of $B_{\rm NV}^{\rm bias}=200$~mT, aligned with the [111] direction of the diamond crystal. In this case only one ODMR frequency was measured, $f_1$, and the net magnetic field was defined as $B_{\rm NV}=f_1/\gamma_{\rm NV}$, ignoring the offset which does not enter the $M_z$ map. The $M_z$ maps obtained with these two bias field conditions are shown in Fig.~\ref{FigSI_bias}. The $M_z$ values appear slightly larger at 200 mT versus 5 mT by 5-10\%, which may be partly explained by the paramagnetic contribution to the magnetization. Despite this small difference, it is clear that 200 mT does not dramatically change the measured magnetization in our ultrathin samples, which remains well short of the magnetization of bulk VI$_3$.   

\begin{figure*}[htb!]
	\begin{center}
		\includegraphics[width=0.5\textwidth]{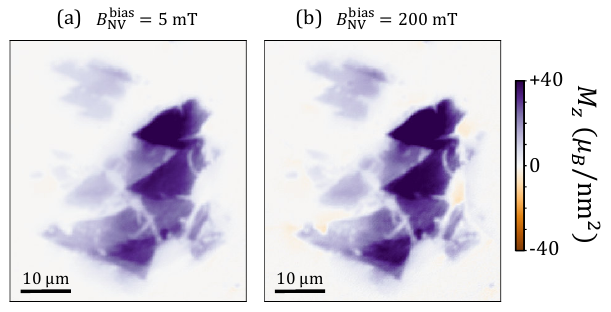}
		\caption{\textbf{Effect of the bias magnetic field.} $M_z$ maps of sample \#1 obtained from the magnetic field maps $B_{\rm NV}$ measured under a bias magnetic field of (a) $B_{\rm NV}^{\rm bias}=5$~mT and (b) $B_{\rm NV}^{\rm bias}=200$~mT. In both cases the bias field is aligned with the [111] direction of the diamond crystal. 
		}
		\label{FigSI_bias}
	\end{center}
\end{figure*}

\section{Ab initio calculations}

In order to test the possibility of a different magnetic behaviour in few-layer VI$_3$ compared to bulk, we performed ab initio calculations of monolayer, bilayer and trilayer VI$_3$ structures. We examined in particular the interlayer exchange coupling, since in a few other van der Waals magnets (e.g. CrI$_3$) it was found to differ between bulk samples (ferromagnetic interlayer coupling, FM) and encapsulated ultrathin samples (antiferromagnetic interayer coupling, AFM) due to a different stacking arrangement~\cite{Li2019,Chen2019}. For VI$_3$, we found that the most stable bilayer is in its AB stacking form with interlayer AFM coupling, while trilayer is also in interlayer AFM state and has an ABA stacking form. This may provide a possible (partial) explanation for the weaker magnetization observed experimentally in few-layer samples compared to bulk VI$_3$. Moreover, we discovered a magnetic second-order phase transition when the van der Waals interlayer distance changes, not only in bilayers (FM) but also in trilayers (AFM). Consistent with the bulk electronic properties, monolayers and bilayers should be Mott insulators with calculated band gaps in the range of 0.80-0.90 eV, while trilayers interestingly become half metals whether in ABA or ABC stacking forms. The methods and results are presented in detail below.

\subsection{Methods}

The structure searching is performed by using USPEX \cite{Glass2006,Oganov2006} combined with Vienna ab initio simulation package (VASP) \cite{Kresse1993,Kresse1996} in the projector augmented wave (PAW) method \cite{Kresse1999}. The exchange-correlation energy was treated within the generalized gradient approximation (GGA) using the PBE functional \cite{Perdew1996} with the van der Waals (vdW) correction proposed by Grimme \cite{Grimme2006} under zero damping DFT-D3 method. Brillouin Zone (BZ) integrations were carried out using $\Gamma$-centered sampling grids with resolution of $2\pi\times0.04$~\AA~for structure optimizations. The criteria of energy and atomic force convergence are set to $10^{-7}$~eV and $10^{-3}$~eV/\AA. The initial magnetic moment for V atom is set to $2~\mu_B$/V with the easy $z$-axis. It is well-known that first principles calculations usually need to add U values for many transition metal compounds, here we take ${\rm U} = 3.25$ eV for V element, by considering some earlier instructions \cite{Ong2013}. We set up a 25~\AA~large vacuum size for avoiding the interlayer interactions between different periodic unit cells for all 2D structures. In order to carefully consider the interlayer distances, we built a post-process work flow based on pymatgen \cite{Ong2013} to check its final stacking structures with energy as fitness.

\subsection{Results}

\begin{figure*}[t!]
	\begin{center}
		\includegraphics[width=0.9\textwidth]{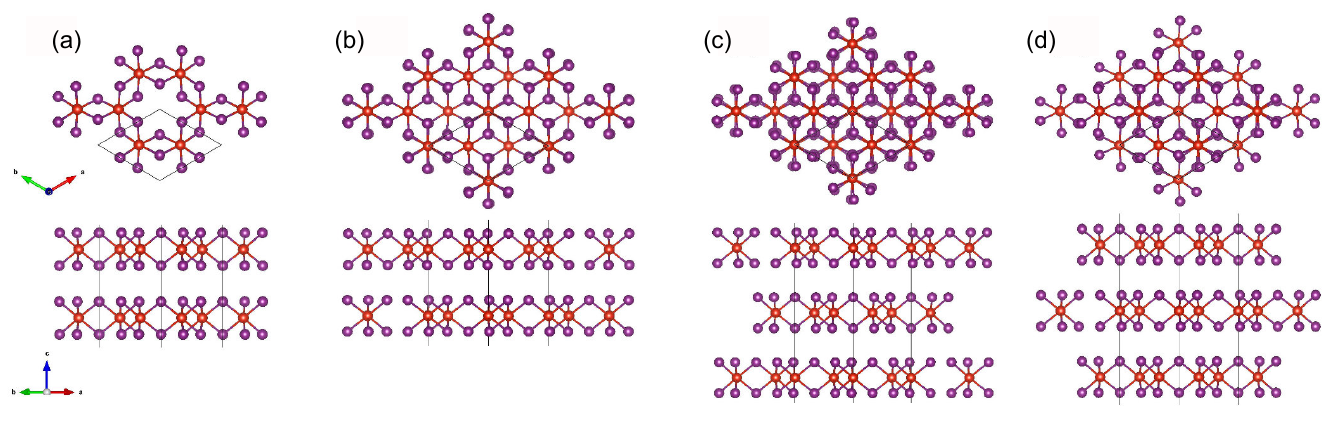}
		\caption{2D structures of VI$_3$ under different stacking forms ($2\times2$ cells): top and side views of (a) AA bilayer; (b) AB bilayer; (c) ABC trilayer; (d) ABA trilayer. 
		}
		\label{FigSI_Zhenhai_structures}
	\end{center}
\end{figure*}

\begin{table*}[tb!]
	\begin{center}
		\includegraphics[width=0.5\textwidth]{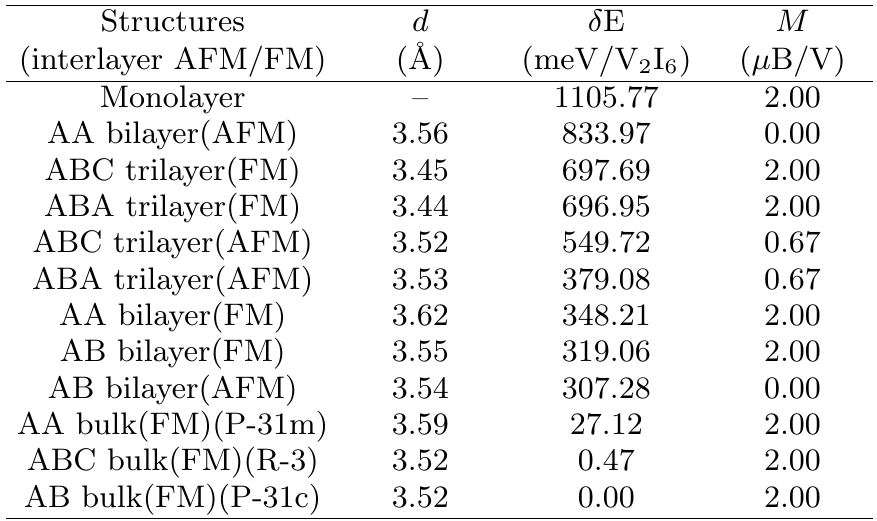}
		\caption{The vdW layer distances ($d$), formation energy differences ($\delta E$) and total magnetic moments ($M$) for 2D and bulk VI$_3$ structures, obtained from GGA-PBE+U calculations. 
		}
		\label{FigSI_Zhenhai_Table}
	\end{center}
\end{table*}

\begin{figure*}[t!]
	\begin{center}
		\includegraphics[width=1\textwidth]{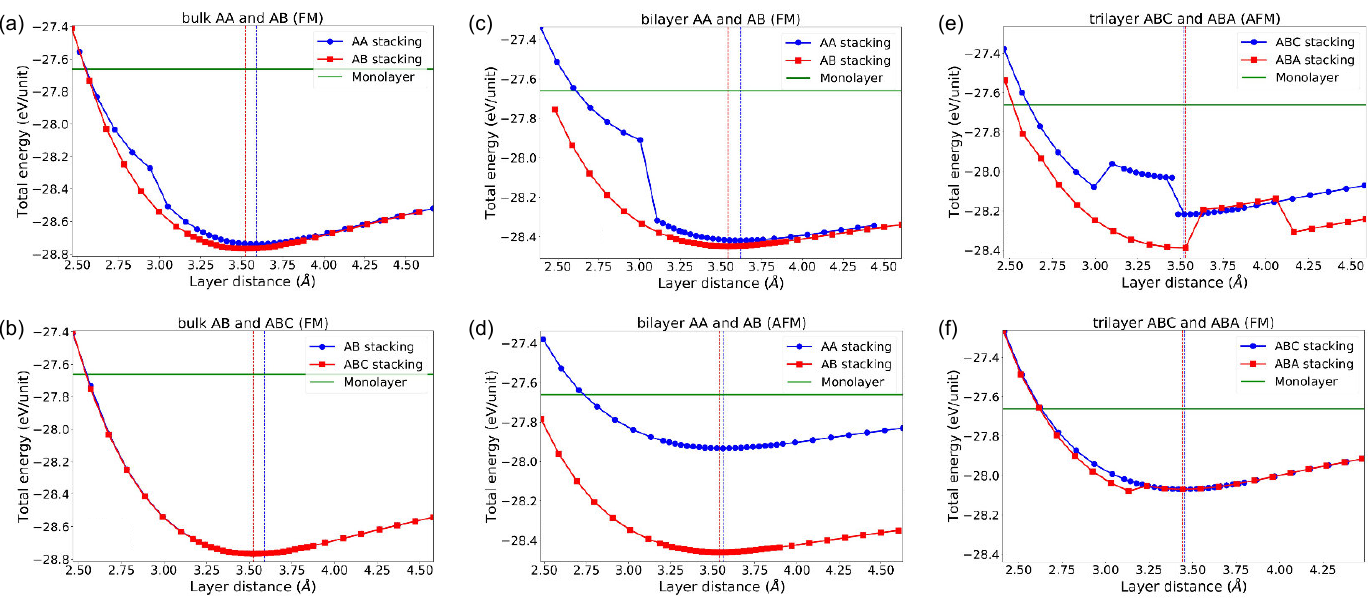}
		\caption{Total energies under variable vdW distances: (a) bulk AA and AB stacking under interlayer ferromagnetic (FM) states; (b) bulk AB and ABC stacking under interlayer FM states; (c) bilayer AA and AB stacking under interlayer antiferromagnetic (AFM) states; (d) bilayer AA and AB stacking under interlayer FM states; (e) trilayer ABC and ABA stacking under interlayer AFM states; (f) trilayer ABC and ABA stacking under interlayer FM states; The vertical dashed lines present the optimized vdW layered distances and the horizontal line means the energy position of monolayer under intralayer FM state, here the U value considered as 3.25 eV in all GGA-PBE+U calculations. 
		}
		\label{FigSI_Zhenhai_stability}
	\end{center}
\end{figure*}

Starting from ferromagnetic VI$_3$ bulk and monolayer, we have investigated four more 2D structures by GGA-PBE+U spin polarized calculations, as shown in Fig.~\ref{FigSI_Zhenhai_structures}. Based on the work flow, we firstly checked the bulk final stacking structures. As listed in listed in Table~\ref{FigSI_Zhenhai_Table}, the lowest formation energy is obtained for the P-31c structure (AB stacking), closely followed by the R-3 structure (ABC stacking), both with FM interlayer state. This is consistent with experiments, where the P-31c structure was observed at room temperature and the R-3 below 79 K~\cite{Son2019,Tian2019,Kong2019}. The variation of the formation energy with the vdW distance between layers is shown in Fig.~\ref{FigSI_Zhenhai_stability}a,b giving 3.52~\AA~with the ABC stacking bulk, which is comparable with the experiments (3.43~\AA) and means the DFT-D3 vdW correction method is reasonable. In Table~\ref{FigSI_Zhenhai_Table}, we can notice that the monolayer has the highest formation energy, as large as 1105.77 meV/V$_2$I$_6$ (or 138.22 meV/atom) above the most stable AB stacking bulk. Meanwhile, bilayers and trilayers are much more energetically favourable, which means they are easier to be achieved in experiments. 

Then, we checked the corresponding multilayers and find that the stable bilayer is in its AB stacking form with interlayer AFM state, while the most stable trilayer is also in interlayer AFM state and has an ABA stacking form, as shown in Fig.~\ref{FigSI_Zhenhai_stability}c-f. More interestingly, here we discovered a magnetic second-order phase transition when the vdWs interlayer distance changes, not only in interlayer FM bilayers but also in interlayer AFM trilayers. Its theoretical explanation can be done by the crystal field theory (CFT), related with the different V occupied magnetic orbitals. Considering AB stacking bilayer, there are two different V atomic positions, V1 and V2 have the same nearest neighbours environment, while V3 and V4 have another one. In our DFT simulations, the V1$/$V2 occupied positions have larger magnetic moments than the V3$/$V4 positions, which implies the splitting of degenerate orbitals under the framework of CFT. For trilayers, although the ABA and ABC stacking forms have different phase transition vs vdW distance positions, as shown in Fig.~\ref{FigSI_Zhenhai_stability}e, they both prefer to have AFM interlayer coupling states, with ABA stacking trilayer the most stable. Therefore, interlayer AFM bilayers can contribute nothing and trilayers have only $0.67~\mu_B$ per V atom ($\mu_B$/V, which is identical to $\mu_B$/f.u.) for magnetization. 

The prevalence of AFM interlayer coupling in few-layer VI$_3$, as identified by our calculations, provides a possible explanation to the weak magnetization observed in the experiments reported above, of $\approx0.4~\mu_B$/f.u. on average for flakes up to 9 layer thick. One could imagine, for instance, that a certain number of layers (increasing with thickness) are always magnetized in the opposite direction to the majority, thus reducing the net magnetization by a proportionality factor. In the bilayer case, a mixture of AFM/FM interlayer coupling caused by local stacking faults could explain why the measured magnetization (at the scale of our spatial resolution) is weak but non-zero. However, we stress that the data can also be explained by assuming FM interlayer coupling together with an intrinsically weak per-layer magnetization of $\approx0.4~\mu_B$/f.u. The reason for such a weak magnetization could be related to disorder caused by degradation of our ultrathin samples, but could also be due to an intrinsic mechanism yet to be identified. We note that the measured magnetization of bulk VI$_3$ of $\approx1.0~\mu_B$/f.u.~\cite{Son2019,Kong2019} is already significantly weaker (by a factor 2) than predicted by ab initio calculations. This indicates that the magnetic behaviour of VI$_3$ is currently not fully captured by ab initio calculations. Future work aiming to correlate the magnetization with the stacking arrangement could shed light onto this system.   

\begin{figure*}[tb!]
	\begin{center}
		\includegraphics[width=0.8\textwidth]{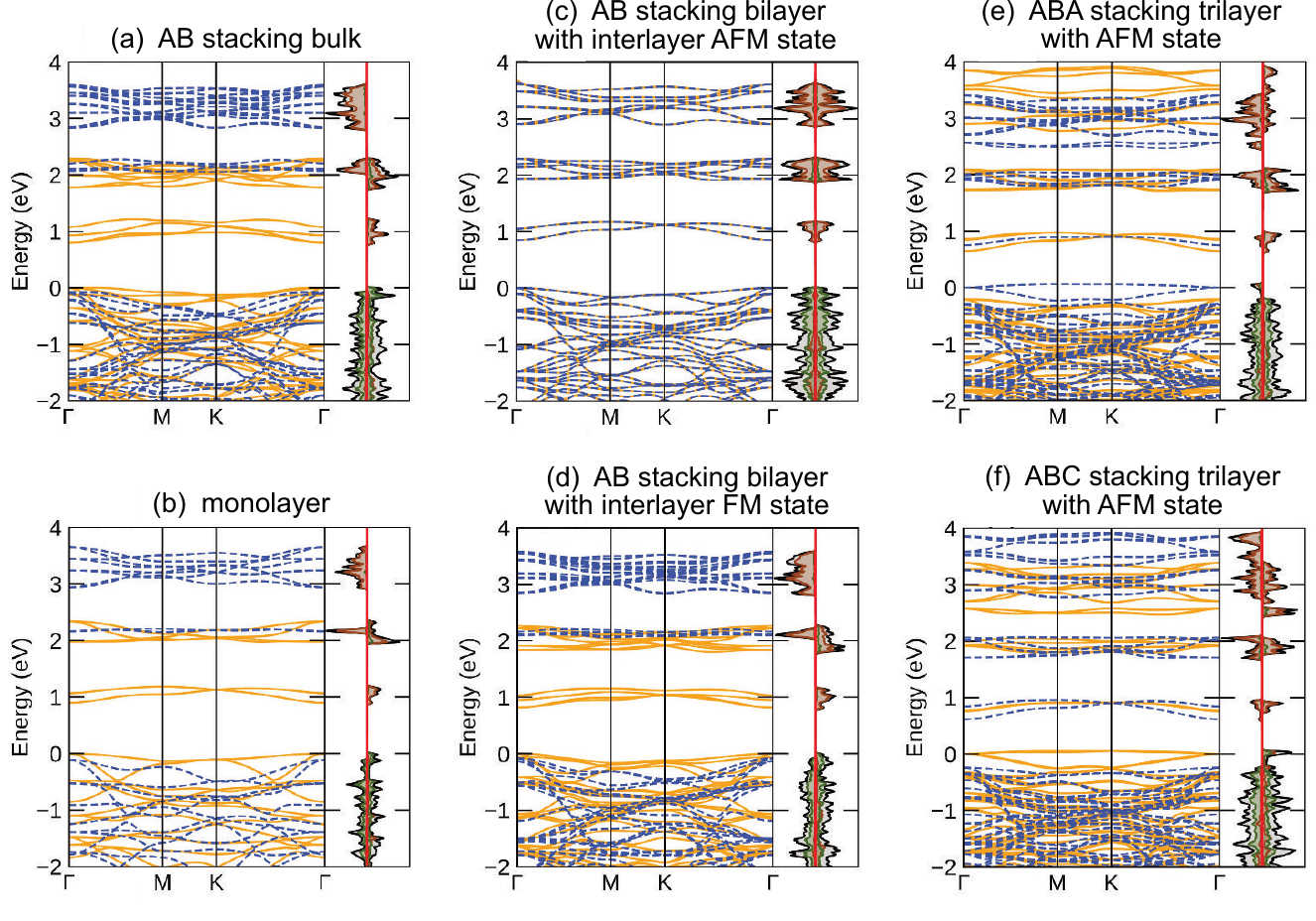}
		\caption{The electronic band structures and density of states (DOS) of bulk and 2D VI$_3$: (a) AB stacking bulk; (b) VI$_3$ monolayer; (c) AB stacking bilayer with interlayer AFM state; (d) AB stacking bilayer with interlayer FM state; (e) ABA stacking trilayer with AFM state; (c) ABC stacking trilayer with AFM state; The majority and minority spin bands are plotted in solid orange and dashed blue lines; The total DOS, V($d$) and I($p$) partial DOS are plotted as black, brown and green lines respectively. Fermi level has been set to zero. 
		}
		\label{FigSI_Zhenhai_bands}
	\end{center}
\end{figure*}

\begin{figure*}[tb!]
	\begin{center}
		\includegraphics[width=1\textwidth]{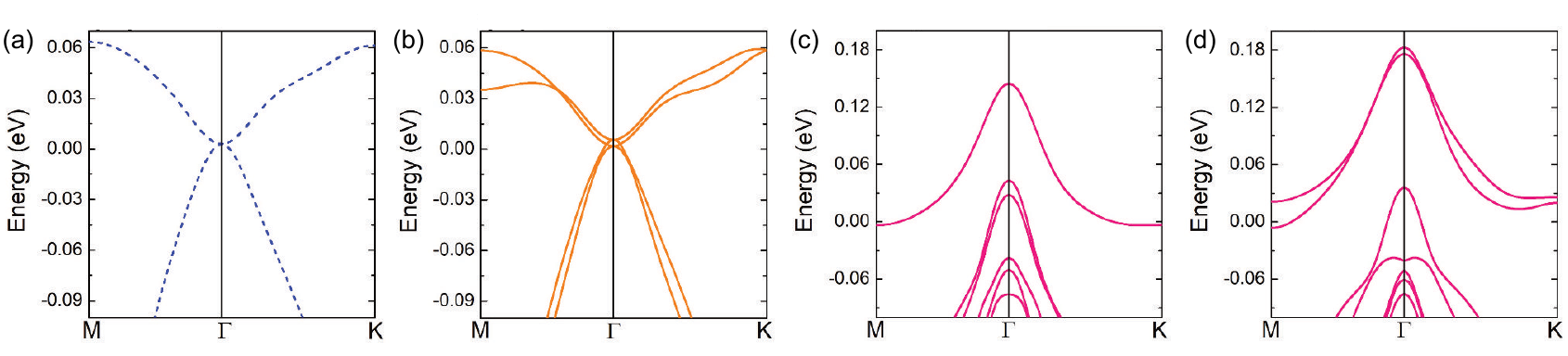}
		\caption{The electronic band structures of VI$_3$ trilayers near Fermi level without and with spin-orbit coupling (SOC): (a) minority spin bands of ABA stacking trilayer; (b) majority spin bands of ABC stacking trilayer; (c) SOC bands of ABA stacking trilayer; (d) SOC bands of ABC stacking trilayer; Fermi level has been set to zero. 
		}
		\label{FigSI_Zhenhai_SOC}
	\end{center}
\end{figure*}

Finally, we theoreticaly investigated the electronic properties of the 2D VI$_3$ structures. Quite different from a previous work on its monolayer and the R3 phase \cite{He2016}, we find that bulk, monolayer and bilayers are all insulators with calculated band gaps in the range of 0.80-0.90 eV, as shown in Fig.~\ref{FigSI_Zhenhai_bands}a-d, which are also consistent with the recent experiment in Ref.~\cite{Son2019}. Interestingly, band structures of trilayers present them as half metals whether in ABA or ABC stacking forms, where the majority spin is insulating with ABA and metallic with ABC stacking, as shown in Fig.~\ref{FigSI_Zhenhai_bands}e-f. In Fig.~\ref{FigSI_Zhenhai_SOC}, we plot the bands near the Fermi level without and with spin-orbit coupling (SOC), and find it can open relative small negative SOC bandgaps at the $\Gamma$ point, which further confirms its metallic behaviour different from monolayer and bilayers.

\end{widetext}

\end{document}